\documentclass[lettersize,journal]{IEEEtran}
\usepackage{amsmath,amsfonts}
\usepackage{algorithmic}
\usepackage{array}
\usepackage[caption=false,font=normalsize,labelfont=sf,textfont=sf]{subfig}
\usepackage{textcomp}
\usepackage{stfloats}
\usepackage{url}
\usepackage{verbatim}
\usepackage{graphicx}
\def\BibTeX{{\rm B\kern-.05em{\sc i\kern-.025em b}\kern-.08em
    T\kern-.1667em\lower.7ex\hbox{E}\kern-.125emX}}
\usepackage{balance}
\usepackage{cite}
\usepackage{amssymb}
\usepackage{color}
\usepackage{xcolor}
\usepackage{hyperref}
\hypersetup{hidelinks=true}
\usepackage{algorithm}
\usepackage{amsthm}
\newtheorem*{problem}{Problem}
\usepackage{booktabs,multirow,makecell}
\newcommand{\best}[1]{\textbf{#1}}
\newcommand{\worst}[1]{\textcolor{red!70!black}{#1}}
\usepackage{comment}

\begin{document}
\title{Untangling Co-Drift: Proactive Multi-Intent Failure Prediction and Root-Cause Disambiguation for Self-Driving Networks}
\author{Md. Kamrul Hossain, Walid Aljoby
\thanks{\textit{(Corresponding author: Walid Aljoby.)}
Md. Kamrul Hossain is with Information and Computer Science Department, King Fahd University of Petroleum and Minerals, Dhahran 31261, Saudi Arabia. \\
Walid Aljoby is with Information and Computer Science Department, and IRC for Intelligent Secure Systems, King Fahd University of Petroleum and Minerals, Dhahran 31261, Saudi Arabia.
}}
\markboth{Journal of \LaTeX\ Class Files,~Vol.~18, No.~9, September~2020}%
{How to Use the IEEEtran \LaTeX \ Templates}
\maketitle

\begin{abstract}
The vision of self-driving networks that monitor, reason, and act upon themselves with minimal human intervention relies on tightly coupled monitoring, analytics, and actuation functions.
In this work, we treat these functions as three operational macro-intents: continuous telemetry (monitoring), real-time analytics (inference and control), and programmatic actuation (rule deployment), and formalize the health of each function as an \emph{intent} that the network must continuously satisfy.
A critical, yet underexplored, challenge stems from the causal coupling among these intents, where a singular fault within one macro-intent propagates as a co-drift and subsequently triggers cascading, symptomatic anomalies across the remaining intents.
This ambiguity makes it exceedingly difficult for existing, reactive approaches to distinguish the true root-cause intent from symptomatic victim intents, and their reliance on threshold-crossing detection leaves insufficient time for proactive remediation. We introduce \textbf{MILD} (Multi-Intent Learning and Disambiguation), a novel framework that reformulates intent assurance from reactive drift detection to proactive failure prediction. Grounded in our three-macro-intent formulation of the self-driving control loop, MILD employs a teacher-augmented Mixture-of-Experts architecture with a hybrid objective that jointly optimizes intent failure prediction and root-cause attribution.
MILD enables KPI-level diagnostics via SHAP explainability and dynamic intent failure urgency estimation via multi-horizon modeling.
Our extensive evaluation of MILD across three environments of increasing realism, from a controlled statistical benchmark, to a microservices application, to an SDN-based edge-to-cloud testbed, demonstrates that MILD achieves high failure detection rates, strong remediation lead times, and accurate intent-level root-cause disambiguation.
This positions MILD as a practical enabler of closed-loop assurance in next-generation autonomous networks.
\end{abstract}

% \section{INTRODUCTION}
% \label{sec:intro}
% \IEEEPARstart{T}{he} cloud-native transformation of enterprise, data-center, and 5G/6G networks has introduced unprecedented operational demands, motivating a shift from imperative configuration to Intent-Based Networking (IBN) \cite{10529727, 10620891, 9925251}. 

\section{INTRODUCTION}
\label{sec:intro}
\IEEEPARstart{M}{odern} networks are increasingly expected to operate as \text{self-driving networks} that continuously monitor, reason about, and act upon themselves in a closed loop, with minimal human intervention~\cite{feamster2017and,princeton, ardestani2025nwdafenabledanalyticsclosedloopautomation}. Realizing this vision in cloud-native enterprise, data-center, and 5G/6G networks has introduced unprecedented operational demands, motivating a shift from imperative configuration to Intent-Based Networking (IBN)~\cite{10529727, 10620891, 9925251}. Following RFC 9315~\cite{rfc9315}, IBN leverages programmable infrastructures~\cite{d2023orchestran, sharma2023comprehensive} to translate high-level declarative goals into verifiable low-level configurations and maintain compliance through continuous, closed-loop assurance. 
A critical component of the IBN paradigm is \textbf{intent assurance}~\cite{10575429, 11334180}, the continuous process of verifying that the network's operational state complies with its intended goals.
%Within 5G and beyond, this capability is increasingly enabled by analytics-driven architectures such as the 3GPP Network Data Analytics Function (NWDAF), which provides the monitoring and analytics needed to support closed-loop automation~\cite{9824403,ardestani2025nwdafenabledanalyticsclosedloopautomation}.
When this compliance is broken, an intent failure starts, leading to a service disruption or an SLA violation. Often, these failures are preceded by \textbf{intent drift}~\cite{rfc9315}, a subtle and gradual deviation of the network's behavior from its intended state that, if unaddressed, culminates in an intent failure.

Solving intent drift raises two challenges that existing work has yet to jointly address. \text{First}, detection must be \emph{proactive}: early drift detection is necessary but insufficient, as what autonomous network management demands is prediction with enough lead time to intervene before a service disruption occurs. \text{Second}, and more critically, modern IBN environments are \textit{multi-intent} systems, in which multiple intents coexist, interact, and share infrastructure. As a result, a fault in one intent can propagate to others and generate ambiguous, cascading symptoms. The observed alert may originate from one intent, while the others are merely victims of the same underlying fault. Current methods struggle to \textit{disambiguate} this cause--victim relation, which limits both attribution and remediation~\cite{11073595, 10575429, 9615580, 10001426}.

The cornerstone of intent drift detection is the real-time analysis of network Key Performance Indicators (KPIs)~\cite{10770652, 10575429}. Anomalous KPI trends, such as sustained increases in latency, drops in throughput, or unusual fluctuations in resource usage, are often the earliest tangible precursors to intent drift~\cite{9925251, 10770652, 10575429}. Yet, current approaches typically treat these KPIs as independent, single-intent signals. They alarm only once a given metric crosses a fixed threshold, well after degradation is already significant~\cite{10575429, 10001426, 11073595}. This single-KPI, single-intent view is precisely why multi-intent ambiguity remains unresolved. Without a model of how faults propagate across coexisting intents, an observer cannot tell whether an anomalous KPI reflects a root cause or a downstream symptom.

This ambiguity is especially consequential when the coexisting intents are not arbitrary tenant services, but the very functions that realize the self-driving loop itself. Indeed, the assurance loop described above is not merely applied \emph{to} the network, it is realized \emph{by} network services. A telemetry pipeline that feeds it observations, an analytics engine that performs its reasoning, and an API/actuation layer that carries out its decisions, each of which is itself governed by an intent. A self-driving network is therefore only as reliable as the intents governing its own monitoring, analytics, and actuation functions. Consequently, a KPI-level fault within one of these functions is not merely an application-level issue but a fault \emph{inside} the assurance loop, one that can silently propagate and compromise the loop's ability to close. To illustrate this recursive dependency, consider the self-driving loop realized as three co-located microservice intents, an API gateway, a telemetry pipeline, and an analytics service, sharing edge resources~\cite{10.1145/3501297, 10.1145/3580305.3599934}. A root-cause CPU fault on the analytics service intent can degrade its throughput and trigger cascading symptoms, such as API latency spikes and telemetry queue backpressure. The resulting KPI traces appear correlated across all intents. This \textit{co-drift} behavior confounds conventional detectors, making it exceedingly difficult to untangle the true root-cause intent from symptomatic victims or estimate failure urgency.

We formalize this recursive structure by casting the self-driving control loop, spanning monitoring, analytics, and actuation, into three representative macro-intents, and build our problem formulation around this abstraction rather than around any single application. A \textbf{Telemetry Intent} governs a data ingestion pipeline responsible for collecting, buffering, and forwarding monitoring data, where `ingestion' refers to this collection and queuing process (monitored by KPIs like \textit{telemetry queue} length, indicating potential backpressure). An \textbf{Analytics Intent} governs a data processing service performing computations on the ingested telemetry (monitored by KPIs like \textit{analytics throughput}). An \textbf{API Intent} governs a request-handling gateway responsible for serving user or service requests with low latency and high availability (monitored by KPIs like \textit{api latency}). These intents map directly onto the self-driving loop's Monitoring, Analytics-and-Control, and Writing-Rules-and-Actions blocks (Sec.~III-A), so that any self-driving network can be treated as an instance of the same three-intent structure. They are monitored via a combination of system-level KPIs (e.g., \textit{CPU\%}, \textit{Mem\%}) and application-specific KPIs (e.g., \textit{api latency}, \textit{telemetry queue}, and \textit{analytics throughput} as detailed in Table~\ref{tab:kpis}). This formulation captures both the shared underlying infrastructure and the distinct intent-level behaviors that must be jointly monitored and disambiguated.

Building on this three-intent abstraction, we introduce \textbf{MILD}, a framework that reformulates intent assurance from reactive drift detection to \textbf{proactive intent failure prediction}. This predictive objective is the key to advancing the state of the art as it compels the model to learn the subtle, \textbf{causal patterns} within ambiguous KPI signals that precede an ultimate failure event. This approach not only provides earlier warnings of impending failures but also enables \textbf{intent-level disambiguation} to identify the true root-cause intent from symptomatic victim intents.

MILD is built around a teacher-augmented Mixture-of-Experts (MoE) architecture~\cite{chen2022towards}, where the gating network is explicitly supervised to resolve root-cause ambiguity under co-drift. A composite loss function combines lead-time-weighted focal loss, knowledge distillation, gate supervision, and expert decorrelation to jointly optimize prediction and attribution.

\begin{figure*}[!t]
    \centering
    \includegraphics[width=0.9\textwidth]{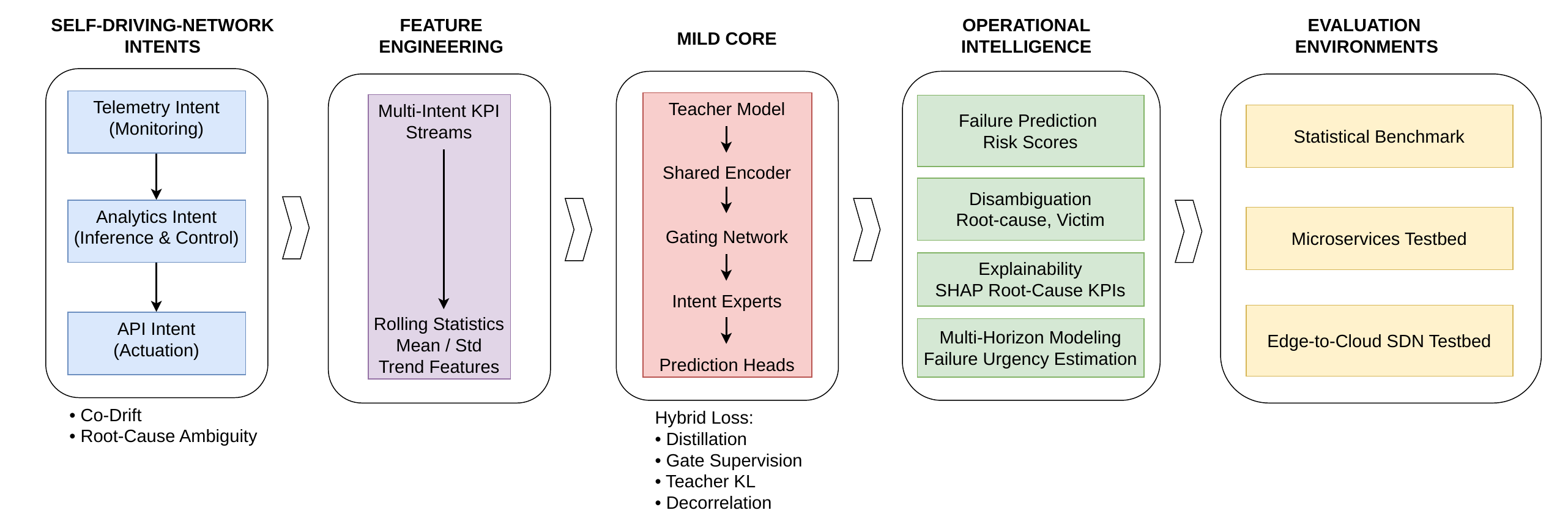}
    \caption{Overview of the proposed framework. KPI streams from the three macro-intents of a self-driving network (Telemetry, Analytics,
and API) are transformed into engineered features and processed by a teacher-augmented MoE architecture. The framework jointly performs proactive failure prediction and root-cause disambiguation through a hybrid optimization objective combining distillation, gate supervision, teacher alignment, and expert decorrelation. Operational intelligence is provided through SHAP-based KPI explanations and multi-horizon time-to-failure estimation.}
    \label{fig:mild_overview}
\end{figure*}

We summarize the main contributions of this work as follows:
\begin{itemize}
    \item We formulate intent assurance around the \textbf{three macro-intents of a self-driving network}—Telemetry, Analytics, and API—and model their causal dependency chain to capture co-drift, where a fault in one intent propagates symptoms to the others. Additionally, we reformulate reactive intent drift detection into a proactive \textbf{multi-intent failure prediction} problem.
    \item We develop a holistic framework, MILD, that operationalizes this predictive formulation, using a specialized MoE architecture to achieve multi-intent failure prediction and \textbf{intent-level disambiguation} in co-drift scenarios.
    \item We extend MILD with a suite of operational intelligence techniques, including post-hoc \textbf{KPI-level disambiguation} via SHapley Additive exPlanations (SHAP)~\cite{10.5555/3295222.3295230} to identify root-cause KPIs and multi-horizon modeling for dynamic intent failure urgency estimation.
    \item We validate MILD on \textbf{three environments}: a controlled statistical benchmark, a containerized microservices testbed, and an SDN-based edge-to-cloud testbed. These datasets explicitly model \textbf{co-drift}, non-linear interactions, and causal ambiguity across intents. We release these datasets to facilitate future research in this domain.
\end{itemize}

\section{Related Work}
The concept of self-driving networks, as an autonomous system combining query-driven measurement, automated inference, and programmatic control, was articulated in early visionary work~\cite{feamster2017and,princeton}. That vision explicitly calls for closing the control loop by coupling real-time telemetry with machine-learning-based inference and SDN-based actuation. We concretize this vision in the IBN context by formalizing the monitoring, analytics/control, and actuation functions of the self-driving loop as operational intents ($\mathcal{I}_\text{tel}$, $\mathcal{I}_\text{anl}$, $\mathcal{I}_\text{api}$) and by framing proactive intent assurance as the core problem that must be solved for the loop to remain closed under faults. While prior IBN assurance work~\cite{11073595,10575429,10770652} addresses individual components of this loop, none models the telemetry, analytics, and actuation functions as a causally coupled system with inherent co-drift dynamics.

Achieving proactive and precise intent assurance that simultaneously predicts failures and disambiguates root causes remains an open challenge in IBN. The core problem is not just detecting intent drift, but predicting a future failure early enough to prevent service impact. While recent work such as NetIntent \cite{11293797} has leveraged Large Language Models (LLMs) for intent realization, its dependence on active tests (e.g., \textit{iperf}, \textit{ping}) and on comparing flow rules between the configuration and operational datastores~\cite{opendaylight2015} of the SDN controller is challenging to scale in complex, large-scale networks.

Another common strategy involves forecasting individual KPIs using time-series models (e.g., LSTMs \cite{10456766, 9615580}) or classifiers (e.g., SVMs \cite{10942926}) and flagging violations against static thresholds. However, these single-KPI forecasts lack the multivariate context necessary to resolve multi-intent ambiguity, where a single fault triggers cascading symptoms across multiple interacting services \cite{10.1145/3580305.3599934, Zanouda_2024}.

Recent works~\cite{hossain2026leaddriftrealtimeexplainableintent,10770652, 10575429} have attempted to formalize the concept of intent drift. The work~\cite{10770652} used an unsupervised method by applying the clustering algorithm DBSCAN to telemetry data to identify anomalous patterns that signify a deviation from normal behavior. While valuable in environments without labeled data, this approach is not trained to recognize the specific multi-variate patterns that are predictive of a future failure and is also prone to false positives from benign network changes. More recent state-of-the-art approaches defined intent drift in terms of KPIs as a deviation from a target state and used LLMs to generate corrective policies~\cite{10575429, 11073595}. While useful, these methods are inherently reactive. They are designed to detect a drift's existence but not its urgency, as they provide no estimate of the time-to-failure. This leaves an unquantified and often insufficient window for proactive remediation. As such, neither of these approaches is explicitly designed to learn the subtle precursors to an impending failure within a specific future window, nor do they offer a solution for the critical task of root-cause disambiguation in multi-intent scenarios.

Recent work~\cite{LIU2026111857} has also explored proactive intent drift prediction at the link level by combining performance forecasting with path similarity, but it remains focused on localized drift detection rather than multi-intent failure disambiguation. Other studies~\cite{DEHGHANBIYAR2025111561, 11334180} address runtime conflict handling in multi-intent settings, yet they do not model early KPI precursors or root-cause ambiguity under co-drift. Monitoring-driven assurance pipelines for application transitions~\cite{VIOLOS2026111872} further show the value of real-time telemetry, but they target application-state changes rather than fixed-horizon failure prediction.

Beyond intent drift, approaches specifically targeting root-cause analysis (RCA) in distributed environments primarily rely on statistical graph traversals and LLMs to isolate system faults. For instance, frameworks like MicroRCA~\cite{9110353} and CauseInfer~\cite{7563819} construct dependency and causality graphs to trace anomalies across services, while advanced agent-based systems like MicroRCA-Agent~\cite{tang2025microrcaagentmicroservicerootcause} and TAMO~\cite{11229957} orchestrate LLMs to fuse multimodal telemetry into comprehensive diagnostic summaries. Despite their localized effectiveness, these methods share a fundamental limitation: they are strictly reactive, initiating computationally heavy or post-hoc diagnostic workflows only after an explicit SLA violation or error has already manifested. Consequently, they struggle to provide actionable early warnings and remain vulnerable to ambiguous, cascading co-drift scenarios.

We bridge these gaps by reformulating the task from reactive drift detection to proactive, fixed-horizon failure prediction. While this predictive technique has been applied to general network outages \cite{Basikolo2023TowardsZD, 9557387}, its application to the specific challenges of IBN assurance has been underexplored. To the best of our knowledge, MILD is the first framework that jointly addresses fixed-horizon intent failure prediction and root-cause disambiguation in multi-intent co-drift scenarios. MILD is explicitly trained to learn the subtle, causal precursors to an impending failure, which enables it to simultaneously achieve long, proactive lead times and, unlike prior methods, provide precise root-cause disambiguation in complex co-drift scenarios.

This manuscript significantly extends our prior work~\cite{hossain2026mildmultiintentlearningdisambiguation}, which introduced the core MILD architecture and presented an initial evaluation using a synthetic benchmark. The present manuscript substantially expands that work in several important directions. First, we broaden its scope by recasting MILD as a general framework for self-driving networks, where the original version represents a particular instance of a more general operational macro-intent and multi-intent system model. Second, we reformulate the problem within the broader context of self-driving networks, providing rigorous new mathematical foundations for the problem space by formalizing three operational macro-intents, intent compliance, intent drift, intent failure, and causal dependencies under co-drift events. Alongside this theoretical expansion, we provide a comprehensive architectural exposition of MILD, detailing the teacher-augmented Mixture-of-Experts design, loss formulations, and gate supervision methodology. Third, to enable evaluation under realistic application and network-level dynamics, we designed two new emulation-based testbeds: a containerized microservices application and an SDN-based edge-to-cloud environment. Finally, we conduct extensive cross-dataset evaluations, robustness analyses, ablation studies, and hyperparameter sensitivity analyses, and we have open-sourced all datasets and experimental artifacts to support full reproducibility.

\section{The MILD Framework}
\label{sec:framework}
\subsection{Motivation: The Three Macro-Intents of a Self-Driving Network}
\label{subsec:macro_intents}
\begin{figure}[!t]
    \centering
    % Ensure you have this figure in the specified path
    \includegraphics[width=0.9\columnwidth]{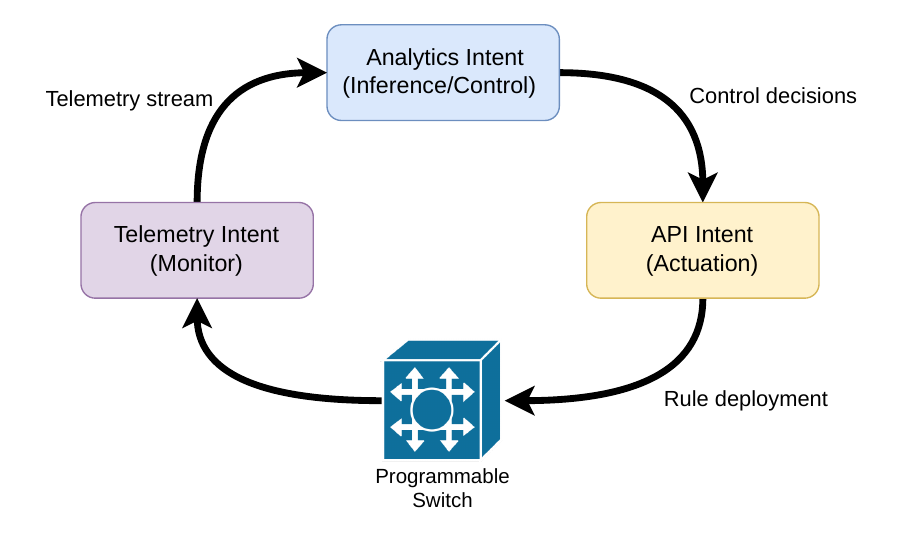}
    \caption{Closed-loop control
architecture for a self-driving network.}
    \label{fig:selfdriving_loop}
\end{figure}
The vision of a self-driving network, as articulated in seminal work on autonomous network management~\cite{feamster2017and,princeton}, centers on a continuous, closed-loop control architecture in which the network monitors itself, reasons about its own state, and takes corrective action with minimal human intervention. In these architectures, the loop is realized through tightly coupled \emph{Monitoring}, \emph{Analytics/Control}, and \emph{Writing Rules and Actions} functions. Building on this architectural view, we cast these functions as three operational \emph{macro-intents} that the network must continuously satisfy. As depicted in Fig.~\ref{fig:selfdriving_loop}, these intents form the functional backbone of the self-driving control loop.
\begin{enumerate}
    \item \textbf{Telemetry Intent ($\mathcal{I}_{\mathrm{tel}}$) — Continuous Monitoring.} The network must perpetually collect fine-grained, query-driven measurements from its own data plane. This intent captures the operator's goal of maintaining an accurate, up-to-date view of system health at all times. It is realized by streaming telemetry pipelines (e.g., in-band network telemetry, Prometheus scraping) and is characterized by KPIs such as ingestion queue length, data-plane throughput, and collection latency. A violation of this intent, for example, a growing telemetry queue that signals backpressure in the data collection pipeline, directly impairs the situational awareness on which all subsequent reasoning depends.
    \item \textbf{Analytics Intent ($\mathcal{I}_{\mathrm{anl}}$) — Real-Time Inference and Control.} The network must continuously process the collected telemetry to infer higher-level properties, detect anomalies, predict failures, and compute corrective actions. This intent corresponds to the \emph{Analytics and Control} block of the self-driving loop and is realized by distributed inference engines and machine-learning models running over streaming data. It is well characterized by KPIs such as analytics throughput and CPU utilization of the inference infrastructure. A violation of this intent, for example, a CPU bottleneck that degrades the processing rate, starves the control plane of the intelligence it needs to act, even when telemetry data is arriving
    correctly.
    \item \textbf{API Intent ($\mathcal{I}_{\mathrm{api}}$) — Rule Writing and Programmatic Action.} The network must expose a reliable, low-latency programmatic interface through which high-level operator policies and automatically derived rules are translated into concrete data-plane configurations. This intent corresponds to the \emph{Writing Rules and Actions} block of the self-driving loop and is realized by SDN controllers, orchestrators, and their northbound APIs. It is characterized by KPIs such as
    end-to-end API request latency and service availability. A violation of this intent prevents corrective actions from being delivered to the network, rendering the control loop open even when the fault has been correctly identified.
\end{enumerate}
These three intents are not merely co-existing workloads; they form a \emph{causal
dependency chain} that is the backbone of any self-driving network. The Telemetry
Intent feeds raw observations to the Analytics Intent; the Analytics Intent derives decisions that are enacted by the API Intent; and the outcome of those actions is, in turn, observed by the Telemetry Intent, closing the loop. This tight coupling means that a fault in any single intent propagates as \emph{cascading, symptomatic anomalies} into the other two, creating a phenomenon we term a \textbf{co-drift}: an ambiguous multi-intent degradation pattern whose root cause is a single, primary intent but whose observable signal is spread across the entire trio. Fig.~\ref{fig:codrift_cascade} illustrates an example of a canonical co-drift event in which a CPU bottleneck in $\mathcal{I}_{\mathrm{anl}}$ (the root cause) causes queue build-up in $\mathcal{I}_{\mathrm{tel}}$ and latency spikes in $\mathcal{I}_{\mathrm{api}}$ (the symptomatic victims).
It is precisely this co-drift ambiguity that makes self-driving network assurance
fundamentally harder than single-intent monitoring. Reactive approaches that alarm only when a KPI crosses a threshold cannot distinguish whether a telemetry queue spike is a primary fault or a victim of an analytics bottleneck. The problem, therefore, requires a framework that can (i)~\emph{proactively predict} an impending failure before it becomes critical, and (ii)~\emph{precisely disambiguate} the true root-cause intent from its symptomatic victims. The proposed MILD framework is designed to solve exactly this dual problem.

\subsection{Problem Formulation}
\label{subsec:formulation}
We now formally define the \emph{Multi-Intent Failure Prediction and Disambiguation} problem for a self-driving network governed by the three macro-intents introduced above.

\subsubsection{Notation}
\label{subsubsec:notation}
The key mathematical symbols used throughout our formulation are summarized in
Table~\ref{tab:notation}, where we describe each symbol explicitly for clarity.
\begin{table*}[!t]
\centering
\caption{Summary of Key Notations}
\label{tab:notation}
\renewcommand{\arraystretch}{1.2}
\scriptsize
\begin{tabular}{@{}p{2.5cm}p{13cm}@{}}
\toprule
\textbf{Symbol} & \textbf{Description} \\ \midrule
\multicolumn{2}{@{}l}{\textit{System and Intent Model}} \\
$t \in \mathbb{Z}_{>0}$ & Discrete time step index (one step = one minute in our setting). \\
$K \in \mathbb{Z}_{>0}$ & Total number of monitored intents ($K=3$ in the concrete use case: Telemetry, Analytics, API). \\
$\mathcal{I} = \{1,\dots,K\}$ & Set of intent indices. \\
$i^*\in\mathcal{I}$ & Index of the true root-cause intent in a co-drift event. \\
$\mathcal{V}\subset\mathcal{I}$ & Set of victim (symptomatic) intent indices, $i^* \notin \mathcal{V}$. \\
$D \in \mathbb{Z}_{>0}$ & Number of raw KPI signals (input dimensionality before feature engineering). \\
$\mathbf{k}_t \in \mathbb{R}^{D}$ & Raw KPI measurement vector at time step $t$. \\
$t_{\mathrm{fail},i}$ & The future time at which intent $i$ will fail if no corrective action is taken. \\
$H \in \mathbb{Z}_{>0}$ & Prediction horizon, in minutes: the maximum advance warning window. \\[4pt]
\multicolumn{2}{@{}l}{\textit{Feature Engineering}} \\
$D' \ge D$ & Dimensionality of the engineered feature space (includes rolling statistics). \\
$\mathbf{x}_t \in \mathbb{R}^{D'}$ & Engineered feature vector at time $t$: $\mathbf{x}_t = \phi_{\mathrm{eng}}(\{\mathbf{k}_{t'}\}_{t' \le t})$. \\[4pt]
\multicolumn{2}{@{}l}{\textit{Teacher Model}} \\
$f^{(T)}$ & A simpler, pre-trained teacher model used for knowledge distillation. \\
$\mathbf{p}^{(T)}_t \in [0,1]^K$ & Vector of per-intent risk scores produced by the teacher model at time $t$; $p^{(T)}_{t,i}$ is the teacher's estimated probability that intent $i$ is approaching failure. \\[4pt]
\multicolumn{2}{@{}l}{\textit{MILD Model Outputs}} \\
$f_\theta$ & The parameterized MILD model with trainable parameters $\theta$. \\
$\mathbf{p}_t \in [0,1]^K$ & Vector of per-intent risk scores at time $t$; $p_{t,i} \in [0,1]$ is the probability that intent $i$ will fail within the next $H$ minutes. \\
$\mathbf{g}_t \in \Delta^{K-1}$ & Root-cause probability distribution at time $t$, lying on the $(K-1)$-simplex ($\sum_{i=1}^K g_{t,i} = 1$, $g_{t,i} \ge 0$); $g_{t,i}$ is the model's confidence that intent $i$ is the underlying root cause of any observed co-drift. \\[4pt]
\multicolumn{2}{@{}l}{\textit{Ground-Truth Labels}} \\
$y^{\mathrm{bin}}_{t,i} \in \{0,1\}$ & Binary failure label for intent $i$ at time $t$: equals 1 iff $t$ falls within the prediction window preceding a failure, i.e., $t_{\mathrm{fail},i} - H \le t < t_{\mathrm{fail},i}$. \\
$y^{\mathrm{ttf}}_{t,i} \ge 0$ & Continuous time-to-failure label for intent $i$ at time $t$: the remaining minutes until failure, zeroed outside the prediction window. Used to weight the loss function so that samples closer to failure are penalized more heavily. \\
$y^{\mathrm{cause}}_{t,i} \in \{0,1\}$ & Root-cause indicator label for intent $i$ at time $t$: equals 1 iff intent $i$ is the annotated root cause \emph{and} $t$ falls within the prediction window. Used exclusively to supervise the gating network. \\[4pt]
\multicolumn{2}{@{}l}{\textit{Loss Function and Optimization}} \\
$\mathcal{L}_{\mathrm{total}}$ & Total composite training loss, minimized end-to-end. \\
$\mathcal{L}_{\mathrm{head},i}$ & Per-head loss for intent $i$, combining focal and distillation terms. \\
$\mathcal{L}_{\mathrm{gate}}$ & Gate loss that trains the gating network for root-cause disambiguation. \\
$\mathcal{L}_{\mathrm{decorr}}$ & Decorrelation regularization loss that enforces expert diversity. \\
$\alpha \in [0,1]$ & Mixing coefficient between focal loss and distillation loss in $\mathcal{L}_{\mathrm{head},i}$. \\
$\lambda_{\mathrm{gate}},\lambda_{\mathrm{decorr}}$ & Scalar weights controlling the relative contribution of $\mathcal{L}_{\mathrm{gate}}$ and $\mathcal{L}_{\mathrm{decorr}}$ in $\mathcal{L}_{\mathrm{total}}$. \\
$w_c, w_T, \lambda_s$ & Hyperparameters of $\mathcal{L}_{\mathrm{gate}}$: weight on the ground-truth cause term, weight on the teacher-alignment term, and sparsity regularization coefficient, respectively. \\[4pt]
\multicolumn{2}{@{}l}{\textit{Alerting}} \\
$\tilde{p}_{t,i}$ & EWMA-smoothed risk score for intent $i$ at time $t$. \\
$\tau_i \in [0,1]$ & Alert threshold for intent $i$: an alert is issued when $\tilde{p}_{t,i} > \tau_i$. \\
$\Gamma_{\max} \ge 0$ & Maximum tolerated false positive rate (alerts per day), used as a constraint during threshold tuning. \\
\bottomrule
\end{tabular}
\end{table*}
\subsubsection{System Model}
\label{subsubsec:system_model}
Consider a self-driving network whose closed-loop operation is sustained by a set of $K$ distinct operational intents,
\begin{equation}
    \mathcal{I} = \{\mathcal{I}_1, \mathcal{I}_2, \dots, \mathcal{I}_K\},
    \label{eq:intent_set}
\end{equation}
where, concretely and without loss of generality, $K = 3$ with
$\mathcal{I}_1 \equiv \mathcal{I}_{\mathrm{tel}}$,
$\mathcal{I}_2 \equiv \mathcal{I}_{\mathrm{anl}}$, and
$\mathcal{I}_3 \equiv \mathcal{I}_{\mathrm{api}}$ as defined in
Sec.III-\ref{subsec:macro_intents}. The health of these intents is jointly observed through
a multivariate time series of $D$ raw Key Performance Indicators (KPIs),
$\{\mathbf{k}_t\}_{t=1}^{\infty}$, where $\mathbf{k}_t \in \mathbb{R}^D$ is the
vector of all KPI measurements collected at discrete time step $t$ (e.g., one sample
per minute). Each KPI $k_{t,d}$ may carry information relevant to one or more intents
simultaneously, and this shared observability is both what makes co-drift diagnostics
possible and what makes them ambiguous.
\paragraph{Intent Compliance and Drift.}
Each intent $\mathcal{I}_i$ is associated with a \emph{compliance region}
$\mathcal{R}_i \subset \mathbb{R}^D$ defined over the KPI space: the network is
\emph{compliant} with intent $i$ at time $t$ if and only if $\mathbf{k}_t \in
\mathcal{R}_i$. An \emph{intent drift} for intent $i$ is a sustained, progressive
deviation of the observed KPIs away from $\mathcal{R}_i$, i.e., a trajectory
$\{\mathbf{k}_t\}$ that is moving toward the boundary of $\mathcal{R}_i$ and, if left
unaddressed, will cross it at some future time $t_{\mathrm{fail},i}$, constituting an
\emph{intent failure}. Formally,
\begin{equation}
    \text{Failure of } \mathcal{I}_i \text{ at } t_{\mathrm{fail},i}:
    \quad \mathbf{k}_{t_{\mathrm{fail},i}} \notin \mathcal{R}_i,
    \quad \mathbf{k}_t \in \mathcal{R}_i \; \forall\, t < t_{\mathrm{fail},i}.
\end{equation}
\paragraph{Causal Dependency and Co-Drift.}
\begin{figure}[!t]
\centering
\includegraphics[width=0.9\columnwidth]{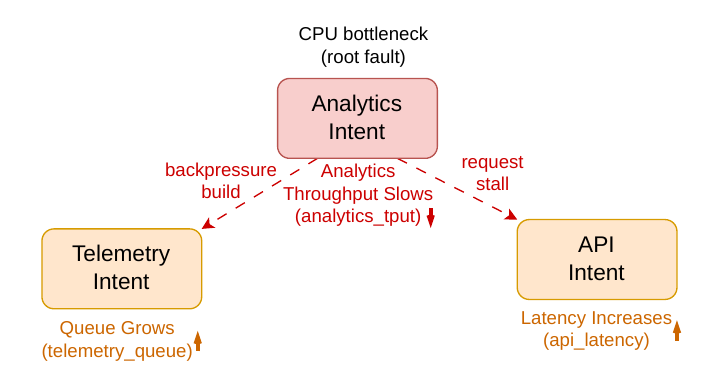}
\caption{Example of a canonical co-drift cascade. A CPU bottleneck in
$\mathcal{I}_\text{anl}$ (root cause, red) propagates via causal
dependency edges (dashed) to $\mathcal{I}_\text{tel}$ and
$\mathcal{I}_\text{api}$ (symptomatic victims, orange), creating
ambiguous, simultaneous KPI anomalies across all three intents.}
\label{fig:codrift_cascade}
\end{figure}
The three macro-intents are not independent. Let
$\mathcal{D} \subseteq \mathcal{I} \times \mathcal{I}$ denote the set of directed
causal dependency edges, where $(i, j) \in \mathcal{D}$ means that a failure of intent
$i$ can induce a symptomatic drift in intent $j$ (e.g.,
$(\mathcal{I}_{\mathrm{anl}}, \mathcal{I}_{\mathrm{tel}})$ and
$(\mathcal{I}_{\mathrm{anl}}, \mathcal{I}_{\mathrm{api}})$ in the cascade illustrated
in Fig.~\ref{fig:codrift_cascade}). A \textbf{co-drift event} occurs when a single
primary fault in the \emph{root-cause intent} $i^* \in \mathcal{I}$ propagates along
the dependency edges in $\mathcal{D}$, causing one or more \emph{victim intents} $\mathcal{V} \subset \mathcal{I} \setminus \{i^*\}$ to simultaneously exhibit anomalous KPI trajectories. In such a scenario, the KPI signal is confounded: each victim intent
$j \in \mathcal{V}$ will show drifting metrics not because of an intrinsic fault of its own, but as a downstream consequence of $i^*$'s degradation. This confounding is the defining challenge of multi-intent assurance. Crucially, the structural topology of
$\mathcal{D}$ is neither assumed to be known a priori nor provided as an explicit input to our model. Instead, MILD implicitly learns to disentangle these latent cause-victim dependencies purely from data, leveraging historical event logs where co-drift instances are labeled with their true root-cause intents.

\subsubsection{Problem Setting: Multi-Intent Failure Prediction and Disambiguation}
\label{subsubsec:problem_statement}
Given the system model above, we formally state the problem that MILD is designed to
solve.
\begin{problem}[Multi-Intent Failure Prediction and Disambiguation]
\label{prob:main}
Let $\mathcal{I} = \{1,\dots,K\}$ be the set of operational intents of a self-driving
network, jointly observed through the multivariate KPI time series $\{\mathbf{k}_t\}$.
The \textbf{Multi-Intent Failure Prediction and Disambiguation} problem requires learning
a function $f_\theta$ that, at each time step $t$ and given the history of KPI
observations up to $t$, simultaneously solves the following two sub-problems:
\end{problem}
\textit{a) Sub-Problem 1: Proactive Multi-Intent Failure Prediction:} For each intent $i\in\mathcal{I}$, produce a scalar risk score $p_{t,i}\in[0,1]$ that remains near 0 during nominal operation ($t\ll t_{fail,i}$) but increases monotonically to cross an alert threshold $\tau_{i}$ at least $H$ minutes before failure. The prediction must minimize false positives by distinguishing genuine pre-failure drifts from benign transient anomalies.
\textit{b) Sub-Problem 2: Root-Cause Disambiguation:} Concurrently, at any time $t$ during a pre-failure drift event, produce a decisive probability distribution $g_{t}\in\Delta^{K-1}$ over the intent set $\mathcal{I}$. The mass of $g_{t}$ must concentrate on the true root-cause intent $i^{*}$ even when symptomatic victims $\mathcal{V}$ dominate the observable KPI signals, providing clear attribution at the exact moment of the first alert ($p_{t,i^*}$ first exceeds $\tau_{i^*}$).
\paragraph{Joint Objective} The two sub-problems must be solved \emph{simultaneously} by a single model, because the disambiguation of $i^*$ is inherently coupled to the prediction of risk scores. Understanding \emph{which} intent is the root cause directly informs \emph{how strongly} each intent should be alarmed, and vice versa. A unified architecture that jointly reasons about prediction and attribution is therefore fundamentally necessary.
\paragraph{Operational Intelligence.}
Beyond the core dual objective, an operationally useful solution must additionally:
\begin{enumerate}
    \item[(a)] Provide \emph{KPI-level explanations} for each alert, identifying which specific metrics drove the risk score, to guide human investigation.
    \item[(b)] Offer \emph{dynamic intent failure urgency estimation} that enables operators to gauge the urgency of an impending failure as it approaches.
    \item[(c)] Operate within a \emph{false positive budget} $\Gamma_{\max}$, ensuring that the alerting system remains trustworthy in production.
\end{enumerate}
The MILD framework, presented in full in the remainder of this section, is the first, to the best of our knowledge, to holistically address this joint problem in the context of
Intent-Based Networking.
\subsubsection{Fixed-Horizon Labeling Strategy}
\label{subsubsec:labeling}
To supervise $f_{\theta}$, we leverage historical event logs to generate three complementary labels at each time step $t$ for each intent $i \in \mathcal{I}$:
\begin{itemize}
    \item \textbf{Binary Failure Label ($y_{t,i}^{bin} \in \{0,1\}$):} Targets failure prediction within a horizon $H$:
    \begin{equation}
        y^{\mathrm{bin}}_{t,i} = \mathbf{1}\big\{t_{\mathrm{fail},i} - H \le t < t_{\mathrm{fail},i}\big\}.
        \label{eq:bin_label}
    \end{equation}

    \item \textbf{Time-to-Failure (TTF) Label ($y_{t,i}^{ttf} \ge 0$):} Weights the loss function to prioritize early failure precursors:
    \begin{equation}
        y^{\mathrm{ttf}}_{t,i} = (t_{\mathrm{fail},i}-t) \cdot y^{\mathrm{bin}}_{t,i}.
        \label{eq:ttf_label}
    \end{equation}

    \item \textbf{Root-Cause Label ($y_{t,i}^{cause} \in \{0,1\}$):} Supervises the gating network specifically for root-cause disambiguation:
    \begin{equation}
    \begin{aligned}
    y^{\mathrm{cause}}_{t,i}
    &=
    \mathbf{1}\big\{
    i = i^*
    \big\}
    \cdot y^{\mathrm{bin}}_{t,i}.
    \end{aligned}
    \label{eq:cause_label}
    \end{equation}
\end{itemize}
Importantly, in a co-drift event caused by intent $i^*$, we assign $y_{t,i^*}^{cause}=1$. For all symptomatic victim intents $j \in \mathcal{V}$, $y_{t,j}^{cause}=0$ despite their binary labels $y_{t,j}^{bin}$ being active.
\subsubsection{Predictive Learning Formulation}
\label{subsubsec:learning_formulation}
To solve the joint problem stated in Sec.III-B-\ref{subsubsec:problem_statement}, we reformulate it as a supervised learning task. The raw KPI vector $\mathbf{k}_t$ is first enriched by a feature engineering function $\phi_{\mathrm{eng}}$ that computes rolling-window statistics (mean and standard deviation over windows of 5 and 15 minutes) across all base KPIs, yielding the enriched feature vector
$\mathbf{x}_t = \phi_{\mathrm{eng}}(\{\mathbf{k}_{t'}\}_{t' \le t}) \in \mathbb{R}^{D'}$ with $D' \gg D$. This enrichment is critical since the rolling statistics encode the \emph{trend} and \emph{volatility} of each KPI that are more diagnostic of pre-failure drift than instantaneous values alone.
A pre-trained teacher model $f^{(T)}$, which is a simpler, independently trained classifier operating on the same features, produces its own per-intent risk score vector $\mathbf{p}^{(T)}_t = f^{(T)}(\mathbf{x}_t) \in [0,1]^K$. These teacher predictions serve two roles: (i) as a soft distillation target for the prediction heads, providing a smooth, calibrated training signal; and (ii) as an auxiliary input to the gating network, supplying a coarse first hypothesis about which intents are at risk, which the gate can then refine.
The MILD model $f_\theta$ is then defined as the parameterized function:
\begin{equation}
    f_\theta \;:\; \big(\mathbf{x}_t,\; \mathbf{p}^{(T)}_t\big) \;\longmapsto\; \big(\mathbf{p}_t,\; \mathbf{g}_t\big),
    \label{eq:mild_mapping}
\end{equation}
where:
\begin{itemize}
    \item $\mathbf{p}_t = \{p_{t,i}\}_{i=1}^K \in [0,1]^K$ is the vector of
    per-intent risk scores that addresses Sub-Problem 1 (Proactive Prediction). Each
    element $p_{t,i}$ is the probability, as estimated by the model, that intent $i$
    will fail within the next $H$ minutes.
    \item $\mathbf{g}_t = \{g_{t,i}\}_{i=1}^K \in \Delta^{K-1}$ is the root-cause
    probability distribution that addresses Sub-Problem 2 (Disambiguation). The
    element $g_{t,i}$ represents the model's confidence that intent $i$ is the
    underlying primary cause of any ongoing system-wide degradation, with
    $\sum_{i=1}^K g_{t,i} = 1$.
\end{itemize}
The architecture of $f_\theta$ that realizes this joint mapping is the
teacher-augmented MoE model described in the following section.
\subsection{MILD Model Architecture}
\label{subsec:architecture}
\begin{figure*}[t!]
    \centering
    \includegraphics[width=0.9\linewidth]{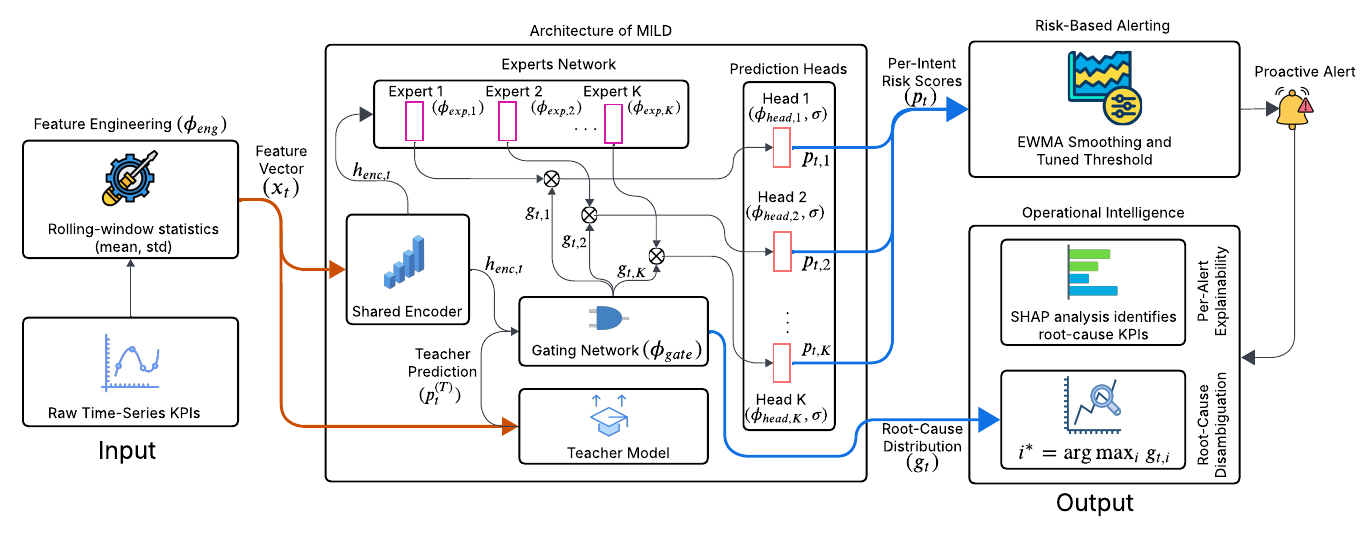}
    \caption{MILD: teacher-augmented MoE with risk prediction, root-cause disambiguation, and per-alert intelligence.}
    \label{fig:mild_framework}
\end{figure*}
Fig.~\ref{fig:mild_framework} shows MILD's flow from KPI features and teacher priors to per-intent risks, root-cause gating, and EWMA-threshold alerting with SHAP explanations.
The system ingests raw time-series KPIs, which are first transformed into an enriched feature vector $\mathbf{x}_t$ by the Feature Engineering module. This vector is then processed in parallel by the pre-trained Teacher Model $f^{(T)}$ and the core MILD model $f_\theta$. The MILD model itself is a teacher-augmented MoE architecture. A Shared Encoder generates a latent system representation $\mathbf{h}_{\mathrm{enc},t}$, which is fed to both a Gating Network and $K$ specialized Expert networks. The Gating Network, conditioned on both $\mathbf{h}_{\mathrm{enc},t}$ and the teacher's predictions $\mathbf{p}^{(T)}_t$, produces the final \textbf{Root-Cause Distribution ($\mathbf{g}_t$)}. It also generates per-expert modulation weights $(g_{t,i})$ that scale each expert's output before it is passed to a corresponding prediction Head. The Heads produce the final \textbf{Per-Intent Risk Scores ($\mathbf{p}_t$)}. These two outputs are then consumed by the operational layers for real-time alerting and root-cause
intelligence, as detailed in Subsec.~\ref{subsec:alerting} and Subsec.~\ref{subsec:intelligence}.
MILD realizes the mapping $f_\theta$ in Eq.~\eqref{eq:mild_mapping} through the
following components:
\begin{itemize}
    \item \textbf{Shared Encoder ($\phi_{\mathrm{enc}}$).} A stack of dense layers maps
    the full engineered feature vector $\mathbf{x}_t$ to a shared latent
    representation:
    \begin{equation}
        \mathbf{h}_{\mathrm{enc},t} = \phi_{\mathrm{enc}}(\mathbf{x}_t) \in \mathbb{R}^{d_h}.
        \label{eq:encoder}
    \end{equation}
    This shared representation encodes the global system state, where $d_h$ is the dimensionality of the shared latent vector, and serves as the common input to both the gating and expert sub-networks, ensuring that all intent-specific branches operate from a consistent, jointly learned view of the network.
    \item \textbf{Gating Network ($\phi_{\mathrm{gate}}$).} This network produces the
    root-cause distribution $\mathbf{g}_t$ by conditioning on both the shared latent
    state and the teacher's coarse predictions. The concatenation
    $[\mathbf{h}_{\mathrm{enc},t};\, \mathbf{p}^{(T)}_t]$ is the input, and a softmax
    output layer guarantees a valid probability distribution:
    \begin{equation}
        \mathbf{g}_t = \mathrm{softmax}\!\Big(\phi_{\mathrm{gate}}\big([\mathbf{h}_{\mathrm{enc},t};\, \mathbf{p}^{(T)}_t]\big)\Big).
        \label{eq:gate}
    \end{equation}
    Including $\mathbf{p}^{(T)}_t$ in the gate's input is the key \emph{teacher
    augmentation}: the teacher's simpler per-intent risk estimates provide an initial soft prior over which intents are at risk, which the gate can confirm or override based on the richer latent representation $\mathbf{h}_{\mathrm{enc},t}$. This allows the gating network to leverage complementary information from two abstraction levels simultaneously, improving disambiguation accuracy especially in the early, ambiguous stages of a co-drift.
    \item \textbf{Experts Network and Prediction Heads ($\phi_{\mathrm{exp},i}$,
    $\phi_{\mathrm{head},i}$).} Each of the $K$ intent-specific experts $\phi_{\mathrm{exp},i}$
    refines the shared latent representation $\mathbf{h}_{\mathrm{enc},t}$ into a
    specialized embedding of dimension $d_e$ that captures the failure precursor patterns unique to
    intent $i$:
    \begin{equation}
        \mathbf{h}_{\mathrm{exp},t,i} = \phi_{\mathrm{exp},i}(\mathbf{h}_{\mathrm{enc},t}) \in \mathbb{R}^{d_e}.
        \label{eq:expert}
    \end{equation}
    The gate score $g_{t,i} \in [0,1]$ then modulates this embedding before it is
    passed to the $i$-th prediction head $\phi_{\mathrm{head},i}$, which maps it to a
    scalar risk score via a sigmoid activation:
    \begin{equation}
        p_{t,i} = \sigma\!\Big(\phi_{\mathrm{head},i}\big(g_{t,i} \cdot \mathbf{h}_{\mathrm{exp},t,i}\big)\Big) \in [0,1].
        \label{eq:head}
    \end{equation}
    This multiplicative modulation introduces a form of soft intent selection: if the gating network assigns low confidence to intent $i$ as the root cause ($g_{t,i} \approx 0$), its expert's embedding is suppressed, reducing its risk score; if intent $i$ is strongly suspected as the root cause ($g_{t,i} \approx 1$), its expert's signal is amplified. This coupling between $\mathbf{g}_t$ and $\mathbf{p}_t$ is what allows the model to jointly reason about prediction and attribution in a single forward pass.
\end{itemize}
\subsection{Hybrid Optimization Framework}
\label{subsec:optimization}
The model is optimized end-to-end by minimizing a composite loss function that combines four distinct objectives, each addressing a specific facet of the joint problem.
\subsubsection{Per-Head Loss: Lead-Time-Weighted Focal Loss and Distillation}
Each prediction head $i$ is trained by minimizing a weighted combination of two terms:
\begin{equation}
    \mathcal{L}_{\mathrm{head},i} = \alpha\, \mathcal{L}_{\mathrm{focal},i} + (1-\alpha)\, \mathcal{L}_{\mathrm{distill},i},
    \label{eq:head_loss}
\end{equation}
where $\alpha \in [0,1]$ controls the balance between the two terms.
$\mathcal{L}_{\mathrm{focal},i}$ is an early-precursor weighted focal loss~\cite{lin2017focal}. The standard focal loss modulates binary cross-entropy by a focusing factor $(1-p_{t,i})^\gamma$, where $\gamma \ge 0$ is a tunable focusing parameter. It down-weights easy (confidently predicted) examples and concentrates learning on hard, ambiguous samples near the decision boundary. To emphasize proactive prediction, this is further multiplied by a temporal weighting factor $w^{\mathrm{ltf}}_{t,i} \propto (y^{\mathrm{ttf}}_{t,i} / H)$, which assigns the highest penalty to samples at the beginning of the prediction window and smoothly decays to a non-zero minimum weight as $t \to t_{\mathrm{fail},i}$. This explicit temporal bias prevents the model from lazily relying on late-stage, obvious failure symptoms, forcing it instead to learn the subtle, early-stage drifts that maximize operational lead time. Additionally, a static class-balancing weight is applied to the active failure windows to compensate for the rarity of failure events relative to nominal network operation.
$\mathcal{L}_{\mathrm{distill},i}$ is a Kullback–Leibler (KL) divergence term~\cite{hinton2015distilling} that penalizes the difference between the model's output distribution and the teacher's softened predictions. Let $\ell^{(T)}_{t,i}=\mathrm{logit}(p^{(T)}_{t,i})$ and
$\ell_{t,i}=\mathrm{logit}(p_{t,i})$ denote the teacher and student logits, respectively. Then:
\begin{equation}
    \mathcal{L}_{\mathrm{distill},i}
    =
    \mathrm{KL}\!\Big(
    \sigma\!\big(\ell^{(T)}_{t,i}/\mathcal{T}\big)
    \;\big\|\;
    \sigma\!\big(\ell_{t,i}/\mathcal{T}\big)
    \Big),
\end{equation}
where $\mathcal{T}>1$ is the distillation temperature, and $\sigma(\cdot)$ denotes the sigmoid activation function. Following standard knowledge distillation practice, temperature scaling is applied to the teacher and student logits before the sigmoid transformation, producing softened Bernoulli distributions for KL matching. This distillation guides the student model to match the teacher's generalization behavior, reducing overfitting to the hard binary labels and improving robustness to unseen drift patterns.
\subsubsection{Gate (Disambiguation) Loss}
The gating network is trained specifically for root-cause disambiguation via three
complementary objectives:
\begin{equation}
\begin{aligned}
\mathcal{L}_{\mathrm{gate}} =\;& w_c\, \mathrm{KL}\!\big(\mathbf{y}^{\mathrm{cause}}_t \,\|\, \mathbf{g}_t\big) + w_T\,\mathrm{KL}\!\big(\mathbf{d}^{(T)}_t \,\|\, \mathbf{g}_t\big)
\\
&+ \lambda_s \sum_{i=1}^K g_{t,i}(1 - g_{t,i}),
\end{aligned}
\label{eq:gate_loss}
\end{equation}
where $\mathbf{y}^{\mathrm{cause}}_t = \{y^{\mathrm{cause}}_{t,i}\}_{i=1}^K$ is the vector of root-cause labels and $\mathbf d_t^{(T)}$ denotes the teacher distribution obtained by applying temperature-scaled softmax to the teacher logits. For the KL divergence computation, $\mathbf{y}^{\mathrm{cause}}_t$ is $L_1$-normalized to form a valid probability distribution. To prevent arbitrary routing during nominal operation, the ground-truth cause alignment term, $KL(y_{t}^{cause}||g_{t})$, is explicitly masked (computed as zero) unless an active root-cause label is present in the prediction window. Conversely, the teacher alignment term, $KL(d_{t}^{(T)}||g_{t})$, remains unmasked at all times. This design choice allows the teacher's distribution to act as a continuous regularizer during nominal operation, gently guiding the gating network toward a stable, uncommitted state when no faults are present. We empirically found that this mechanism improves overall disambiguation performance and prevents premature overconfidence in the gating network.
The three terms serve distinct roles: (i) the first term aligns $\mathbf{g}_t$ with the ground-truth root-cause labels $\mathbf{y}^{\mathrm{cause}}_t$, directly
supervising the disambiguation objective; (ii) the second term aligns $\mathbf g_t$ with the teacher-derived intent
distribution $\mathbf d_t^{(T)}$, allowing the teacher's simpler analysis to regularize the gate's behavior and improve generalization; (iii) the third term is a sparsity regularizer that penalizes high-entropy, non-decisive gate outputs—it is minimized when $g_{t,i} \in \{0, 1\}$, encouraging the gate to make confident, single-intent attributions rather than spreading probability mass across all intents ambiguously.
\subsubsection{Head Decorrelation Loss}
To promote functional diversity among the $K$ experts and prevent them from collapsing to a single, shared representation (which would negate the MoE design philosophy), we apply a DeCov regularization term~\cite{cogswell2016reducing} that penalizes the off-diagonal elements of the empirical covariance matrix $\widehat{\mathbf{\Sigma}} \in \mathbb{R}^{K d_{\mathrm{head}} \times K d_{\mathrm{head}}}$ of the concatenated head feature vectors across a training batch, where $d_{\mathrm{head}}$ is the hidden dimensionality of each prediction head:
\begin{equation}
    \mathcal{L}_{\mathrm{decorr}} = \big\|\, \widehat{\mathbf{\Sigma}} - \mathrm{diag}(\widehat{\mathbf{\Sigma}}) \,\big\|_F^2.
    \label{eq:decorr_loss}
\end{equation}
Minimizing this term drives each expert's learned representation to be uncorrelated with the others, ensuring that each head captures intent-specific failure precursor patterns that are orthogonal to those captured by its peers.
\subsubsection{Total Training Objective}
The four components are combined into a single composite loss function:
\begin{equation}
    \mathcal{L}_{\mathrm{total}} = \sum_{i=1}^K \mathcal{L}_{\mathrm{head},i}
    + \lambda_{\mathrm{gate}}\, \mathcal{L}_{\mathrm{gate}}
    + \lambda_{\mathrm{decorr}}\, \mathcal{L}_{\mathrm{decorr}},
    \label{eq:total_loss}
\end{equation}
where the hyperparameters $\lambda_{\mathrm{gate}} \ge 0$ and
$\lambda_{\mathrm{decorr}} \ge 0$ govern the relative importance of the gate and
decorrelation objectives with respect to the primary prediction objective. The full set
of hyperparameters $\{\alpha, \lambda_{\mathrm{gate}}, \lambda_{\mathrm{decorr}}, w_c,
w_T, \lambda_s, \mathcal{T}\}$ is tuned on a held-out validation set.
\subsection{Failure Prediction and Alerting}
\label{subsec:alerting}
The learned per-intent risk scores $p_{t,i}$ are translated into operational alerts
through a two-stage policy. First, the raw score is smoothed using an Exponential
Moving Average (EWMA)~\cite{Hyndman_Athanasopoulos_2021} with span $W_i$ to suppress
high-frequency noise and stabilize predictions:
\begin{equation}
    \tilde{p}_{t,i} = \mathrm{EWMA}(p_{t,i};\, W_i).
\end{equation}
An alert for intent $i$ is then issued at time $t$ if the smoothed score exceeds the
tuned threshold:
\begin{equation}
    \text{Alert for } \mathcal{I}_i \text{ at } t \;\Longleftrightarrow\; \tilde{p}_{t,i} > \tau_i.
\end{equation}
The EWMA span $W_i$ and threshold $\tau_i$ are tuned jointly via grid search on the validation set to maximize expected lead time subject to a false positive budget:
\begin{align}
    \max_{\tau_i,\, W_i} \quad & \mathbb{E}\big[\mathrm{LeadTime}(\tau_i, W_i)\big] \label{eq:opt_leadtime} \\
    \text{s.t.} \quad & \mathbb{E}\big[\mathrm{FalsePositivesPerDay}(\tau_i, W_i)\big] \le \Gamma_{\max}. \label{eq:opt_constraint}
\end{align}
This constrained formulation ensures that the alerting policy is operationally viable. Maximizing warning time is meaningful only if the false alarm rate is controlled below the operator's tolerance $\Gamma_{\max}$.
\subsection{Disambiguation and Operational Intelligence}
\label{subsec:intelligence}
\subsubsection{Intent-Level Disambiguation}
At the time of an alert for any intent $i$, the gating network's output $\mathbf{g}_t$ immediately provides the root-cause attribution required by Sub-Problem 2. Because the gate has been explicitly trained against the $y^{\mathrm{cause}}$ labels via
$\mathcal{L}_{\mathrm{gate}}$, its output is a calibrated probability distribution over the intents that reflects the model's causal understanding of the system, not merely a correlation-based ranking. The most likely root-cause intent is identified as:
\begin{equation}
    i^* = \arg\max_{i \in \mathcal{I}} \; g_{t,i}.
\end{equation}
This mechanism correctly handles both simple single-drift scenarios where one $g_{t,i}$ is naturally dominant, and the harder co-drift scenarios, where the gate must assign high mass to $i^*$ despite the victims $\mathcal{V}$ exhibiting the most visually conspicuous KPI anomalies.
\subsubsection{KPI-Level Disambiguation via SHAP}
To provide operators with the granular, \emph{KPI-level} explanation required by operational intelligence objective~(a), each alert on the identified root-cause intent $i^*$ is accompanied by a SHAP analysis. For the alert at time $t$, SHAP computes the
contribution $\phi_j$ of each input feature $x_{t,j}$ to the final risk score $p_{t,i^*}$:
\begin{equation}
    p_{t,i^*} \approx \mu + \sum_{j=1}^{D'} \phi_j(x_{t,j}),
\end{equation}
where $\mu$ is a baseline expected value and $\phi_j$ is the Shapley value of feature
$j$, satisfying the efficiency axiom $\sum_j \phi_j = p_{t,i^*} - \mu$.
Features with large positive $\phi_j$ are the KPIs most responsible for the elevated risk score, providing immediate diagnostic guidance: the operator knows not only \emph{which} intent is the root cause, but \emph{which specific metrics} within that intent are driving the impending failure.

\subsubsection{Dynamic Failure Urgency Estimation via Multi-Horizon Models}
\label{dynamic_failure_urgency}
\begin{figure}[!t]
    \centering
    % Ensure you have this figure in the specified path
    \includegraphics[width=0.9\columnwidth]{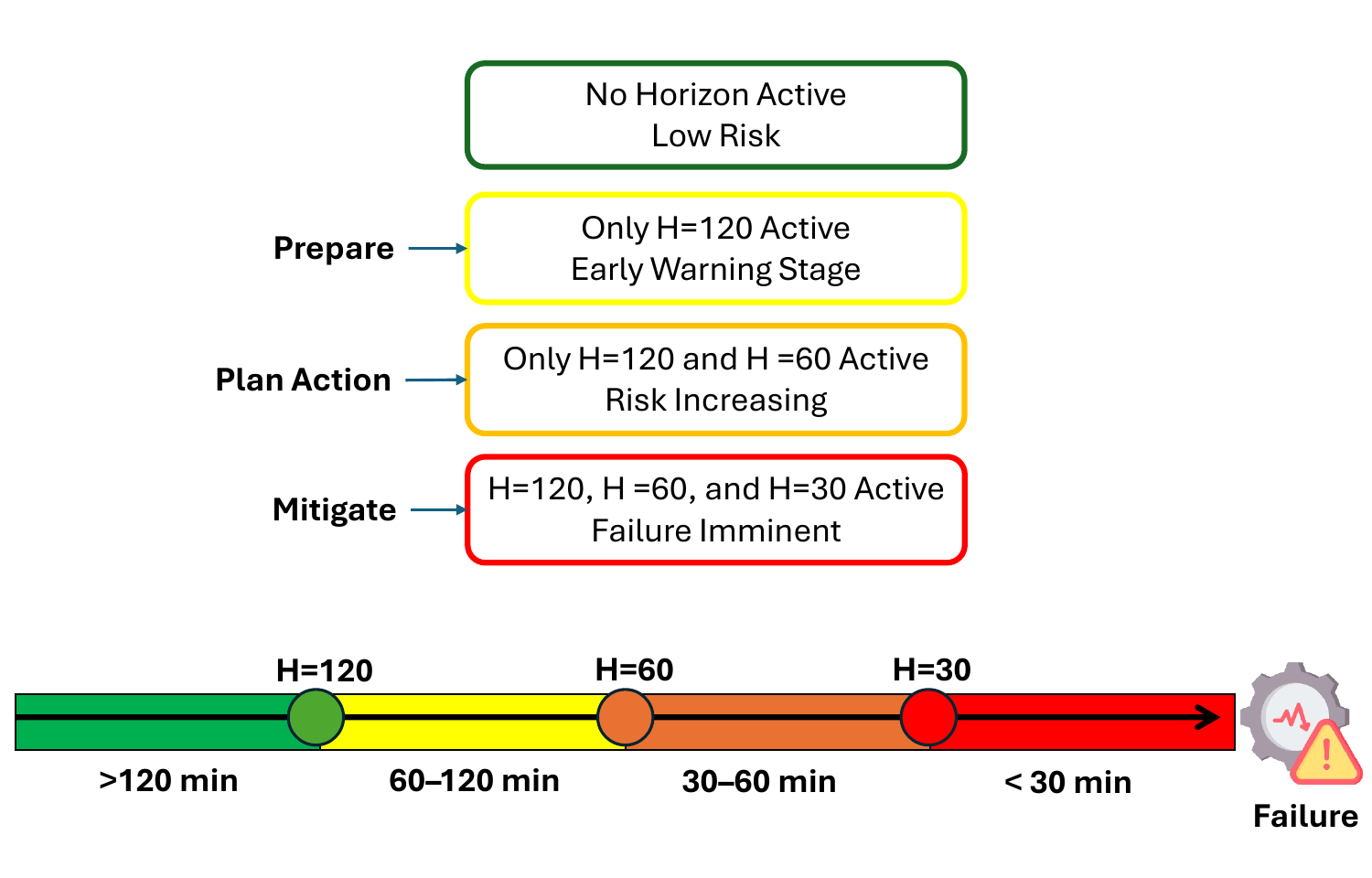}
    \caption{Multi-horizon MILD deployment. Multiple models trained at different prediction horizons (e.g., $H=120$, $H=60$, and $H=30$ minutes) provide progressively refined estimates of failure urgency.}
    \label{fig:multi_horizon}
\end{figure}
To address operational intelligence objective~(b), MILD enables deployment as an ensemble $\{f_{H_1}, f_{H_2}, \dots, f_{H_N}\}$, where each model $f_{H_n}$ is trained with a different prediction horizon $H_n$. At inference time, the combination of alert states across models provides a dynamically tightening bound on the true TTF. Concretely, for a three-model ensemble $\{f_{H_{\mathrm{long}}}, f_{H_{\mathrm{mid}}}, f_{H_{\mathrm{short}}}\}$, if $f_{H_{\mathrm{long}}}$ alerts but $f_{H_{\mathrm{short}}}$ does not, the estimated TTF is bounded as $\text{TTF} \in (H_{\mathrm{short}}, H_{\mathrm{long}}]$, giving operators a concrete window of urgency for scheduling remediation. As the failure approaches and shorter-horizon models begin to alert, this bound narrows, providing a continuously updated countdown of urgency. This is illustrated in Fig.~\ref{fig:multi_horizon} where each model answers a different question. For example, the long-horizon model (trained with $H{=}120$ min) predicts whether a failure is likely within the next two hours; the medium-horizon model ($H{=}60$ min) predicts whether a failure is likely within the next hour; and the short-horizon model ($H{=}30$ min) predicts whether a failure is likely within the next thirty minutes.

\section{Dataset Generation}
\begin{table}[t!]
\centering
\scriptsize
\caption{Base KPIs Used For Dataset Preparation}
\label{tab:kpis}
\begin{tabular}{@{}lp{3cm}l@{}}
\toprule
\textbf{KPI Name} & \textbf{Description} & \textbf{Category} \\ \midrule
\textit{CPU} & CPU utilization percentage & System Resource \\
\textit{Mem} & Memory utilization percentage & System Resource \\
\textit{Disk} & Disk utilization percentage & System Resource \\ \addlinespace
\textit{snet} & Composite network health score & Network Health \\
\textit{sri} & Composite service reliability score & Network Health \\
\textit{telemetry\_queue} & Length of the data ingestion queue & Intent-Specific \\
\textit{analytics\_tput} & Throughput of the data processing service & Intent-Specific \\
\textit{api\_latency} & Latency for API requests & Intent-Specific \\ \bottomrule
\end{tabular}
\end{table}

In this section, we describe how we designed three datasets for the evaluation of MILD. Our datasets not only follow the behavior of real-world network KPIs but also contain precisely labeled, complex failure scenarios that are difficult to isolate in operational data. Such complexity arises frequently in modern distributed systems like cloud-native applications and microservices \cite{10.1145/3501297, 10.1145/3580305.3599934}.

Each dataset begins with the 8 base KPI features shown in Table~\ref{tab:kpis}, representing our raw input dimension $D=8$. While the system resource and intent-specific KPIs are standard metrics, the network-level KPIs (\textit{snet} and \textit{sri}) are bounded composite scores (0--100) whose generation methodology reflects the specific evaluation environment. In the microservices testbed, they are generated using seasonal drift with stochastic variation and injected fault penalties. Conversely, in the SDN-based testbed, they are derived from active network measurements where \textit{snet} summarizes actual packet loss, excess delay, and jitter from active ICMP probes, while \textit{sri} combines the rolling HTTP request success ratio with excess latency and packet loss to characterize end-to-end service reliability.

Within the machine learning pipeline, these base KPIs (raw time-series data) are augmented by computing rolling-window statistics (mean and standard deviation) over 5 and 15-minute windows. In addition, two more features are added, \textit{CPU\_delta} and \textit{sri\_delta}, which are computed as the first-order temporal difference with respect to the features \textit{CPU} and \textit{sri}, respectively. Together, these rolling statistics and temporal differences expand the raw measurements into the enriched input feature space (dimension $D^{\prime}$). Finally, target features (labels) are generated from the annotated events based on the fixed-horizon labeling strategy detailed in Sec. III-B-\ref{subsubsec:labeling}. For each intent, this creates the three distinct labels required for MILD's hybrid loss function: a binary label ($y^{\mathrm{bin}}$), a continuous TTF label ($y^{\mathrm{ttf}}$), and the root-cause label ($y^{\mathrm{cause}}$).

\subsection{Controlled Statistical Benchmark}
To establish a rigorous baseline for MILD, we developed a Python-based algorithmic data generator that produces a highly controlled, multivariate time-series benchmark. Unlike operational data where failure conditions are often obscured, this controlled algorithmic environment allows us to inject precise, mathematically defined failure signatures. The baseline KPIs are modeled to reflect realistic operational patterns, incorporating seasonality for day/night cycles and Gaussian noise for natural system fluctuations. Progressive metric degradation during failure events is modeled via localized random walks. To test model robustness against false positives, the benchmark is heavily injected with unlabeled ``benign mimics'' (i.e., transient anomalies that resolve independently) and random high-load periods. This reflects the challenge of distinguishing true failures in complex systems~\cite{10.1145/3501297}.
The primary feature of this benchmark is the injection of three distinct categories of labeled failure events reflecting known difficulties in real-world network operations \cite{10.1145/3501297, 10.1145/3580305.3599934}:
\begin{itemize}
    \item \textbf{Simple Independent Drifts:} Standard single-intent failures where one KPI degrades.
    \item \textbf{Non-Linear Failures:} Complex single-intent failures where the drift is triggered by subtle, non-linear interactions between multiple KPIs (e.g., XOR-style dependencies). These scenarios represent situations where simple thresholding or linear models often fail because individual metrics might appear normal \cite{10.1145/3501297}.
    \item \textbf{Multi-Intent Co-Drifts:} Scenarios where a single underlying fault creates ambiguous, cascading symptoms across multiple intents, mirroring the challenges of root-cause analysis in distributed microservice systems where failure propagation is common \cite{10.1145/3580305.3599934, 10.1145/3501297}. One intent is explicitly labeled as the \textit{root cause} and another as a \textit{symptomatic victim}.
\end{itemize}
The final benchmark consists of 200,000 minutes of per-minute KPI data, weighted heavily towards complex scenarios to rigorously test the model with 60\% non-linear failures, 20\% multi-intent co-drifts, and 20\% simple independent drifts.
\subsection{Emulation-Based Datasets}
While the statistical benchmark enables controlled evaluation of complex failure patterns, it does not fully capture the dynamics of real-world systems. To complement this, we design two emulation-based testbeds that generate realistic KPI time series data under varying conditions. These environments enable the modeling of multi-intent interactions, network impairments, and service-level dependencies, which are essential for evaluating MILD in practical scenarios. We use the same KPIs as mentioned in Table~\ref{tab:kpis}.
\subsubsection{Microservices Emulation Testbed}
\begin{figure}[!t]
    \centering
    % Ensure you have this figure in the specified path
    \includegraphics[width=\columnwidth]{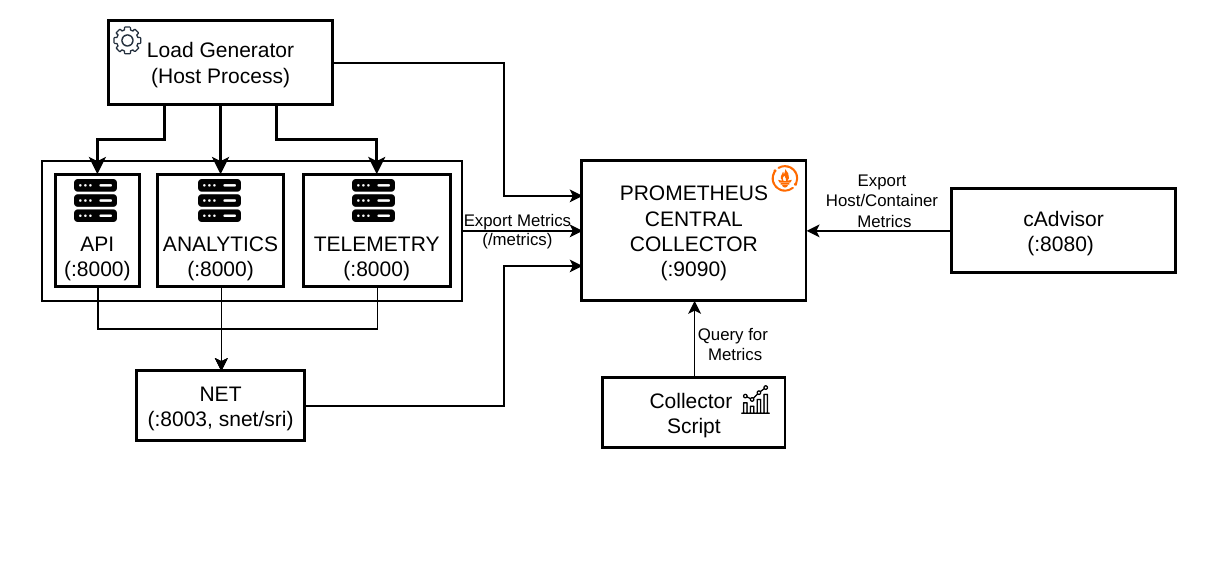}
    \caption{Containerized microservices emulation testbed for KPI generation and monitoring.}
    \label{fig:micro}
\end{figure}
\begin{table}[!t]
\caption{Components and Traffic Profiles of the Microservices Emulation Testbed}
\label{tab:microservices_testbed}
\centering
\footnotesize
\setlength{\tabcolsep}{3pt}
\renewcommand{\arraystretch}{1.15}
\begin{tabular}{>{\raggedright\arraybackslash}p{2.7cm} >{\raggedright\arraybackslash}p{2.3cm} >{\raggedright\arraybackslash}p{3.0cm}}
\toprule
\textbf{Component / Traffic} & \textbf{Nature / Example} & \textbf{Function / Purpose} \\
\midrule
\multicolumn{3}{l}{\textit{System Components}} \\
Load generator & Host process & Simulates user requests \\
API service (8000) & Container service & Processes requests, reports latency \\
Analytics service (8001) & Container service & Processes jobs, reports throughput \\
Telemetry service (8002) & Container service & Ingests events, reports queue size \\
Network KPI service & Container service & Provides network-level KPIs \\
Prometheus & Monitoring server & Scrapes metrics from services \\
cAdvisor & Monitoring exporter & Reports host/container usage \\
Data collection module & Data collector & Queries metrics, builds dataset \\
\midrule
\multicolumn{3}{l}{\textit{Traffic Types}} \\
API request & GET \texttt{/work} & Simulate user requests \\
Analytics job submission & POST \texttt{/submit} & Simulate analytics jobs \\
Telemetry event ingestion & POST \texttt{/ingest} & Simulate event ingestion \\
Metrics scraping & GET \texttt{/metrics} & Collect service KPIs \\
Host/container monitoring & cAdvisor metrics & Collect system KPIs \\
\bottomrule
\end{tabular}
\end{table}
To capture application-level dynamics, we built a containerized microservices testbed (Fig.~\ref{fig:micro}). Three representative services (API, Analytics, and Telemetry) map to latency, throughput, and queue-based intents. A load generator simulates user requests while Prometheus and cAdvisor continuously collect application and system-level KPIs (Table~\ref{tab:microservices_testbed}), yielding multi-variate traces that reflect service behavior and resource contention.

\subsubsection{SDN-based Edge-to-Cloud Emulation Testbed}
\begin{figure}[!t]
    \centering
    % Ensure you have this figure in the specified path
    \includegraphics[width=\columnwidth]{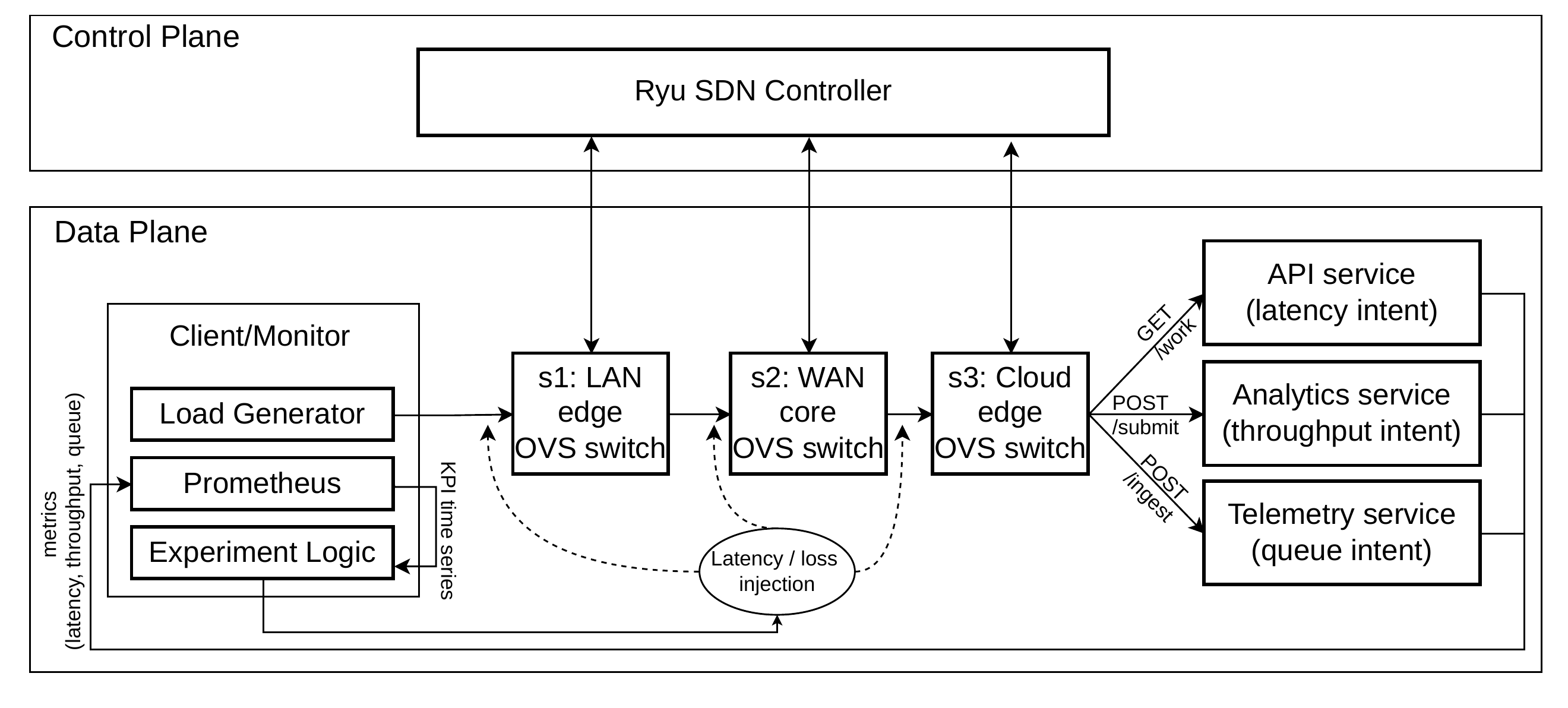}
    \caption{SDN-based edge-to-cloud emulation testbed.}
    \label{fig:Ryu}
\end{figure}

\begin{table}[!t]
\caption{Components and Traffic Profiles of the SDN-based Emulation Testbed}
\label{tab:ryu_testbed}
\centering
\footnotesize
\setlength{\tabcolsep}{3pt}
\renewcommand{\arraystretch}{1.15}
\begin{tabular}{>{\raggedright\arraybackslash}p{2.7cm} >{\raggedright\arraybackslash}p{2.3cm} >{\raggedright\arraybackslash}p{3.0cm}}
\toprule
\textbf{Component / Traffic} & \textbf{Nature / Example} & \textbf{Function / Purpose} \\
\midrule
\multicolumn{3}{l}{\textit{System Components}} \\
Client/monitor host & Host node & Runs load generator, Prometheus, experiment logic \\
API service host & Virtual host/server & Runs API service \\
Analytics service host & Virtual host/server & Runs analytics service \\
Telemetry service host & Virtual host/server & Runs telemetry service \\
LAN edge switch & OVS switch & Connects client to network edge \\
WAN core switch & OVS switch & Represents core network segment \\
Cloud edge switch & OVS switch & Connects services at cloud edge \\
SDN controller (Ryu) & Control plane comp. & Controls switch forwarding via OpenFlow \\
Prometheus & Monitoring server & Scrapes metrics from services \\
Experiment logic & Controller script & Injects faults, collects metrics, writes dataset \\
\midrule
\multicolumn{3}{l}{\textit{Traffic Types}} \\
API workload & GET \texttt{/work} & Simulate user requests \\
Analytics workload & POST \texttt{/submit} & Simulate analytics jobs \\
Telemetry workload & POST \texttt{/ingest} & Simulate telemetry/event ingestion \\
Ping probes & ICMP ping & Measure delay/loss \\
HTTP probes & HTTP requests & Measure reachability \\
Prometheus scrapes & GET \texttt{/metrics} & Collect KPIs \\
OpenFlow control messages & Control traffic & Install/learn forwarding behavior \\
\bottomrule
\end{tabular}
\end{table}
To evaluate network-induced impairments, we implemented a Ryu-controlled SDN testbed (Fig.~\ref{fig:Ryu}) spanning LAN edge, WAN core, and cloud edge segments via Open vSwitch. Application traffic traverses this topology while we inject controlled latency and packet loss into the WAN core (Table~\ref{tab:ryu_testbed}). These network impairments directly impact service-level KPIs, generating complex, fast-propagating co-drift cascades across multiple intents.

\section{Experimental Results}
To ground our experiments, we use the tangible, real-world scenario introduced in Sec.\ref{sec:intro}. Our model is designed to assure the performance of a distributed, cloud-native application by monitoring three representative intents: the API Intent, Telemetry Intent, and Analytics Intent. The health of these intents is monitored using a set of fundamental system-level and application-specific KPIs summarized in Table \ref{tab:kpis}.

\subsection{Experimental Setup}
\label{subsec:MILD_setting}
In this section, we describe the experimental setup used for evaluating MILD across all datasets.
Across all experiments, MILD was trained using the Adam optimizer with a learning rate of $1 \times 10^{-3}$. A blocked cross-validation strategy was employed to prevent temporal leakage, ensuring that each model was trained exclusively on historical data preceding the test block. Training data was further partitioned into an 80/20 split for training and validation. Early stopping was applied with a patience of 8 epochs, monitoring the validation loss.
The core MILD architecture remained identical across all datasets. However, several operational hyperparameters were selected independently for each dataset through validation-based sensitivity analysis to account for differences in temporal scale, failure dynamics, and alerting requirements. These hyperparameters include the prediction horizon $H$, false-positive budget, distillation coefficient ($\alpha$), gate cause-supervision weight ($w_c$), gate teacher-alignment weight ($w_T$), gate sparsity coefficient ($\lambda_s$), batch size, and number of training epochs. The distillation temperature was fixed at $T=2.0$ for all experiments.
The statistical benchmark employed a longer prediction horizon ($H=120$ min), a stricter false-positive budget of 1 alert/day, 10-fold blocked cross-validation (leveraging its extensive data volume), a batch size of 512, and 30 training epochs. The selected MILD hyperparameters were $\alpha=0.9$, $w_c=0.7$, $w_T=0.7$, and $\lambda_s=0.005$.
The microservices emulation testbed dataset employed a shorter prediction horizon ($H=50$ min), a false-positive budget of 3 alerts/day, and 3-fold blocked cross-validation. This adjustment in folds compared to the statistical benchmark was made to ensure sufficient temporal samples remained within each block given the smaller total volume of emulation data. The selected hyperparameters were $\alpha=0.6$, $w_c=0.6$, $w_T=0.3$, and $\lambda_s=0.005$, with a batch size of 128 and 20 training epochs.
The SDN-based edge-to-cloud emulation testbed dataset also used a false-positive budget of 3 alerts/day and 3-fold blocked cross-validation (for the same data-scarcity considerations), but with a shorter prediction horizon ($H=15$ min) that reflects the faster dynamics of network-induced failures. The selected hyperparameters were $\alpha=0.7$, $w_c=0.6$, $w_T=0.3$, and $\lambda_s=0.002$, with a batch size of 256 and 20 training epochs.

\subsection{Baselines}
We compare MILD against five representative baselines trained and evaluated under identical conditions: \textbf{WKPI-Tuned}, a heuristic computing risk scores via pre-trained Logistic Regression coefficients; \textbf{Dist-Target \cite{10575429}}, a target-based method measuring Euclidean distance between current KPIs and a `healthy' target vector; \textbf{LR-OvR \cite{tarekegn2024deeplearningmultilabellearning, 9592636}}, independent One-vs-Rest Logistic Regression classifiers per intent; and \textbf{MLP \& LSTM}, standard supervised neural networks (a multi-layer perceptron and a sequential LSTM) featuring shared encoders and per-intent sigmoid heads.
All baseline outputs are EWMA-smoothed and converted into alerts via thresholding. To ensure a fair and rigorous comparison, all baselines were trained and evaluated under an identical experimental environment as MILD. For reproducibility, all random processes (including data generation, model initialization, and data splitting) were controlled with a global random seed of 42. For compact comparison, detection rates and lead times are reported as macro-averages across intents for all three datasets. We provide the complete implementation of MILD as well as the datasets publicly on GitHub~\cite{HossainMILD}.

\subsection{Results on the Statistical Benchmark}
\label{subsec:results}

\begin{table}[!t]
\centering
\scriptsize
\caption{Performance of MILD on the Statistical Benchmark}
\label{tab:mild_performance_synth}
\begin{tabular}{@{}llc@{}}
\toprule
\textbf{Metric} & \textbf{Details} & \textbf{Value (mean $\pm$ std)} \\ \midrule
\multirow{3}{*}{Failure Detection Rate (\%)} & Analytics & $96.67 \pm 10.00$ \\
 & API & $100.00 \pm 0.00$ \\
 & Telemetry & $100.00 \pm 0.00$ \\ \cmidrule(l){2-3}
\multirow{3}{*}{Avg. Lead Time (min)} & Analytics & $91.53 \pm 7.73$ \\
 & API & $97.63 \pm 12.04$ \\
 & Telemetry & $111.10 \pm 10.07$ \\ \midrule
FP Rate per Day & Overall & $4.97 \pm 6.03$ \\ \midrule
Disambiguation Accuracy (\%) & Root Cause Acc. & $89.67 \pm 10.04$ \\ \bottomrule
\end{tabular}
\end{table}

\begin{table}[t]
\centering
\scriptsize
\caption{Comparison with Baselines on the Statistical Benchmark}
\setlength{\tabcolsep}{2.5pt}
\label{tab:synthetic_compare}
\begin{tabular}{@{}lcccc@{}}
\toprule
\textbf{Model} & \textbf{Detection (\%)} & \textbf{Lead Time (min)} & \textbf{FP/Day} & \textbf{Disamb. Acc. (\%)} \\ \midrule
\textbf{MILD} & 98.89 & 100.09 & 4.97 & 89.67 \\
MLP & 99.44 & 92.77 & 5.05 & 81.03 \\
LSTM & 97.11 & 88.13 & 5.35 & 80.60 \\
WKPI-Tuned & 99.07 & 78.42 & 15.73 & 66.68 \\
Dist-Target & 78.39 & 57.95 & 9.40 & 60.82 \\
LR-OvR & 99.44 & 82.67 & 6.23 & 79.02 \\
\bottomrule
\end{tabular}
\end{table}

Evaluated via 10-fold blocked cross-validation, MILD's performance on the Statistical Benchmark (Table \ref{tab:mild_performance_synth}) demonstrates high reliability, achieving a 98.89\% average Failure Detection Rate. Critically, these predictions provide substantial lead times (averaging 100.09 minutes), offering a substantial operational window for proactive remediation while maintaining a low False Positive Rate (4.97 alerts/day). Furthermore, MILD successfully resolves multi-intent ambiguity by correctly identifying the true root cause in 89.67\% of both co-drift and single-intent failures.

Compared to representative baselines (Table \ref{tab:synthetic_compare}), MILD achieves the most balanced profile. While models like MLP and LR-OvR achieve comparable detection rates, they suffer from significantly shorter lead times and weaker root-cause identification (e.g., MLP: 81.03\%). MILD uniquely combines high detection performance, the longest warning horizon, and superior causal attribution.

\subsection{Results on the Microservices Emulation Testbed}

\begin{table}[t]
\centering
\scriptsize
\caption{Performance of MILD on the Microservices Emulation Testbed}
\label{tab:micro_mild}
\begin{tabular}{@{}llc@{}}
\toprule
\textbf{Metric} & \textbf{Details} & \textbf{Value (mean $\pm$ std)} \\ \midrule
\multirow{3}{*}{Failure Detection Rate (\%)}
 & Analytics & $82.34 \pm 1.03$ \\
 & API & $99.10 \pm 0.62$ \\
 & Telemetry & $92.51 \pm 0.34$ \\ \cmidrule(l){2-3}
\multirow{3}{*}{Avg. Lead Time (min)}
 & Analytics & $29.73 \pm 0.72$ \\
 & API & $34.30 \pm 1.77$ \\
 & Telemetry & $31.25 \pm 0.58$ \\ \midrule
FP Rate per Day & Overall & $8.17 \pm 0.67$ \\ \midrule
Disambiguation Accuracy (\%) & Root Cause Acc. & $66.80 \pm 0.13$ \\ \bottomrule
\end{tabular}
\end{table}

\begin{table}[t]
\centering
\scriptsize
\caption{Comparison with Baselines on the Microservices Emulation Testbed}
\setlength{\tabcolsep}{1.5pt}
\label{tab:micro_compare}
\begin{tabular}{@{}lcccc@{}}
\toprule
\textbf{Model} & \textbf{Detection (\%)} & \textbf{Lead Time (min)} & \textbf{FP/Day} & \textbf{Disamb. Acc. (\%)} \\ \midrule
MILD & 91.32 & 31.76 & 8.17 & 66.80 \\
MLP & 79.77 & 24.00 & 4.41 & 41.22 \\
LSTM & 86.85 & 29.41 & 19.54 & 59.23 \\
WKPI-Tuned & 0.13 & 6.50 & 3.80 & 0.13 \\
Dist-Target & 11.77 & 7.54 & 11.99 & 5.55 \\
LR-OvR & 9.14 & 9.15 & 61.07 & 7.07 \\ \bottomrule
\end{tabular}
\end{table}

Under realistic application-level dynamics (Table \ref{tab:micro_mild}), MILD maintains strong predictive performance, achieving high detection rates for the API (99.10\%) and Telemetry (92.51\%) intents. As expected in noisier environments, the average lead time decreases to 31.76 minutes while remaining sufficient for proactive intervention. Despite complex, non-linear failure interactions, MILD achieves a Root Cause Disambiguation Accuracy of 66.80\% with a controlled false positive rate of 8.17 alerts/day.

Against baselines (Table \ref{tab:micro_compare}), MILD achieves the highest overall detection rate (91.32\%). More importantly, it consistently outperforms the best baseline (LSTM) in disambiguation accuracy by over 7.6 percentage points while more than doubling the performance of the MLP baseline. Baselines that achieve lower false positive rates (e.g., WKPI-Tuned) do so at the cost of severely degraded root-cause attribution, highlighting MILD's superior balance between detection performance, operational reliability, and causal attribution.

\subsection{Results on the SDN-based Emulation Testbed}

\begin{table}[t]
\centering
\scriptsize
\caption{Performance of MILD on the SDN-based Emulation Testbed}
\label{tab:ryu_mild}
\begin{tabular}{@{}llc@{}}
\toprule
\textbf{Metric} & \textbf{Details} & \textbf{Value (mean $\pm$ std)} \\ \midrule
\multirow{3}{*}{Failure Detection Rate (\%)}
 & Analytics & $100.00 \pm 0.00$ \\
 & API & $100.00 \pm 0.00$ \\
 & Telemetry & $76.18 \pm 10.29$ \\ \cmidrule(l){2-3}
\multirow{3}{*}{Avg. Lead Time (min)}
 & Analytics & $10.63 \pm 0.23$ \\
 & API & $7.76 \pm 0.64$ \\
 & Telemetry & $7.48 \pm 1.02$ \\ \midrule
FP Rate per Day & Overall & $8.30 \pm 0.34$ \\ \midrule
Disambiguation Accuracy (\%) & Root Cause Acc. & $88.97 \pm 0.26$ \\ \bottomrule
\end{tabular}
\end{table}

\begin{table}[t]
\centering
\scriptsize
\caption{Comparison with Baselines on the SDN-based Emulation Testbed}
\setlength{\tabcolsep}{1.5pt}
\label{tab:ryu_compare}
\begin{tabular}{@{}lcccc@{}}
\toprule
\textbf{Model} & \textbf{Detection (\%)} & \textbf{Lead Time (min)} & \textbf{FP/Day} & \textbf{Disamb. Acc. (\%)} \\ \midrule
MILD & 92.06 & 8.62 & 8.30 & 88.97 \\
MLP & 55.52 & 8.62 & 5.09 & 51.79 \\
LSTM & 35.91 & 5.97 & 14.73 & 40.77 \\
WKPI-Tuned & 4.34 & 1.26 & 7.20 & 0.51 \\
Dist-Target & 5.26 & 2.12 & 5.14 & 1.54 \\
LR-OvR & 10.56 & 1.15 & 4.70 & 10.26 \\ \bottomrule
\end{tabular}
\end{table}

Under network-level dynamics and impairments (Table \ref{tab:ryu_mild}), MILD achieves perfect detection for the Analytics and API intents and 76.18\% detection for the more challenging Telemetry intent. The tighter prediction horizon and fast failure propagation that is natural to network-driven scenarios result in an average lead time of 8.62 minutes. Despite these challenging network impairments, MILD retains a high Root Cause Accuracy of 88.97\% and a stable false positive rate of 8.30 alerts/day.

Comparison with representative baselines (Table \ref{tab:ryu_compare}) confirms MILD's robustness under network variability. MILD achieves the highest overall detection rate (92.06\%) and substantially outperforms all competing methods in root-cause disambiguation (88.97\% versus 51.79\% for the strongest neural baseline). Competing methods either lose detection capability or fail to provide reliable causal attribution, positioning MILD as a robust framework for dynamic network conditions.

\subsection{Cross-Dataset Robustness Analysis}
\begin{table}[t]
\centering
\scriptsize
\caption{Cross-Dataset Performance Summary of MILD}
\setlength{\tabcolsep}{1.5pt}
\label{tab:cross_dataset}
\begin{tabular}{@{}lcccc@{}}
\toprule
\textbf{Dataset} & \textbf{Detection (\%)} & \textbf{Lead Time (min)} & \textbf{FP/Day} & \textbf{Disamb. Acc. (\%)} \\ \midrule
Statistical & 98.89 & 100.09 & 4.97 & 89.67 \\
Microservices & 91.32 & 31.76 & 8.17 & 66.80 \\
SDN-based & 92.06 & 8.62 & 8.30 & 88.97 \\ \bottomrule
\end{tabular}
\end{table}
Table~\ref{tab:cross_dataset} summarizes MILD's performance across all evaluation environments. For compact comparison, the false positive rates and disambiguation accuracy are reported at the overall level. The results demonstrate that MILD generalizes effectively across datasets with varying levels of realism and complexity. On the statistical benchmark, MILD achieves the highest lead times due to the longer prediction horizon and controlled environment. In contrast, the microservices and SDN-based testbeds introduce realistic noise, service interactions, and network impairments. While this naturally reduces operational lead times, MILD preserves strong early warning capabilities appropriate to each environment, demonstrating its robustness to both application-level and network-level variability.

\subsection{Operational Intelligence and Interpretability}
\subsubsection{Intent-Level Disambiguation}
\begin{figure}[!t]
    \centering
    % Ensure you have this figure in the specified path
    \includegraphics[width=0.8\columnwidth]{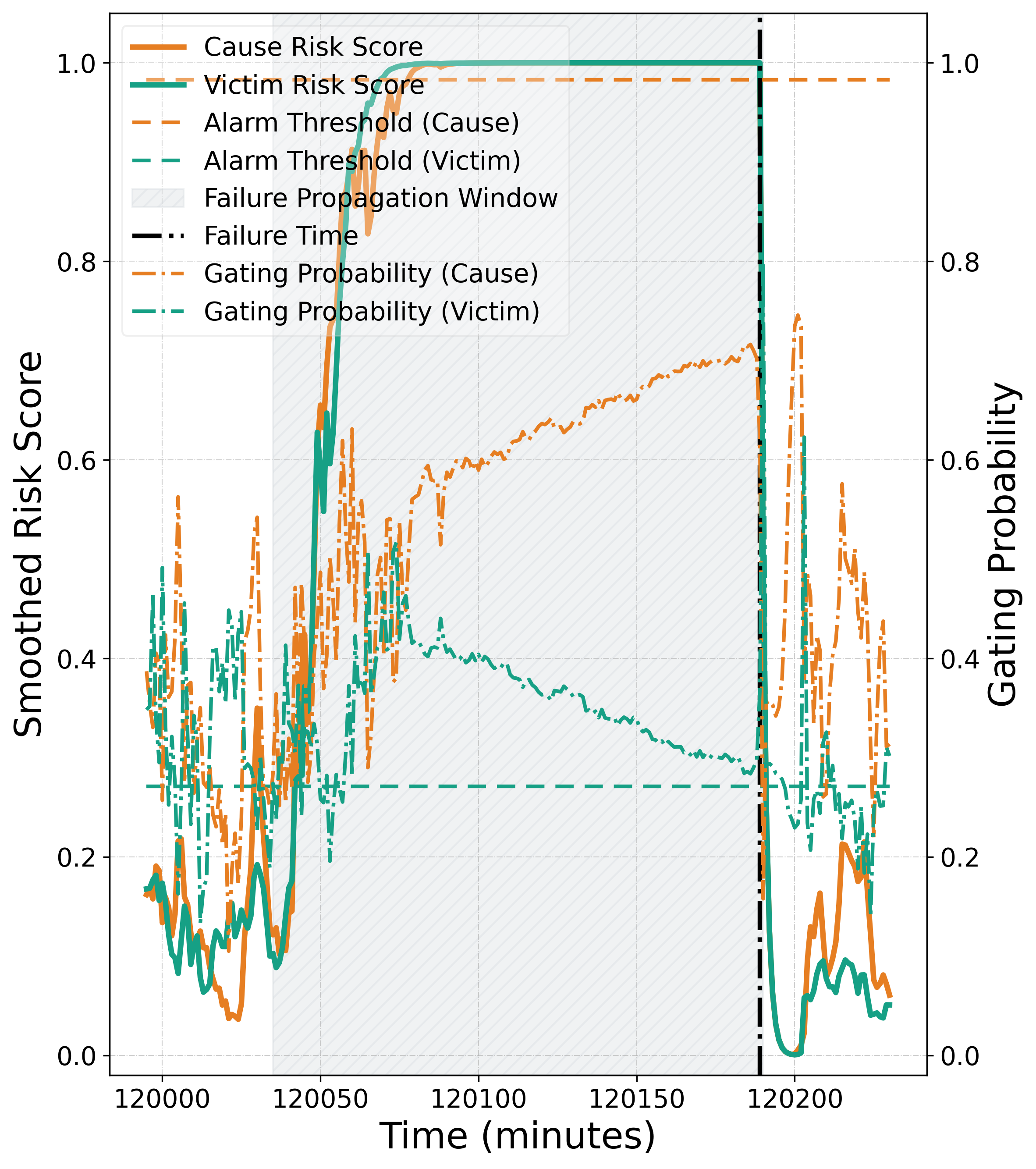}
    \caption{Example of root-cause disambiguation under a co-drift event, showing the evolution of intent risk scores and gating probabilities.}
    \label{fig:disambiguation_example}
\end{figure}
Fig.~\ref{fig:disambiguation_example} illustrates MILD's
disambiguation behavior during a co-drift event in which an
\textit{API} failure acts as the root cause while a
\textit{Telemetry} failure appears as a propagated victim.
Because both intents experience correlated degradation during the
failure propagation window, their smoothed risk scores increase
concurrently and both exceed their respective alarm thresholds.
Such situations are difficult for conventional failure predictors,
which can identify that multiple intents are at risk but cannot
determine which intent initiated the failure.
MILD resolves this ambiguity through its gating network.
Although both intents exhibit high risk scores, the gating
probability assigned to the API intent increases steadily
throughout the drift window, reaching approximately 0.7 before
failure, while the Telemetry gating probability decreases toward
0.3. This indicates that the model attributes the majority of the
observed degradation to the API intent and treats the Telemetry
degradation as a secondary consequence. The resulting separation
between risk estimation and causal attribution enables MILD to
identify the most likely root cause even when multiple intents
simultaneously exhibit elevated risk. This provides operators with a more actionable diagnosis than risk prediction alone.

\subsubsection{KPI-Level Explanation via SHAP}
\begin{figure}[!t]
    \centering
    \includegraphics[width=0.9\columnwidth]{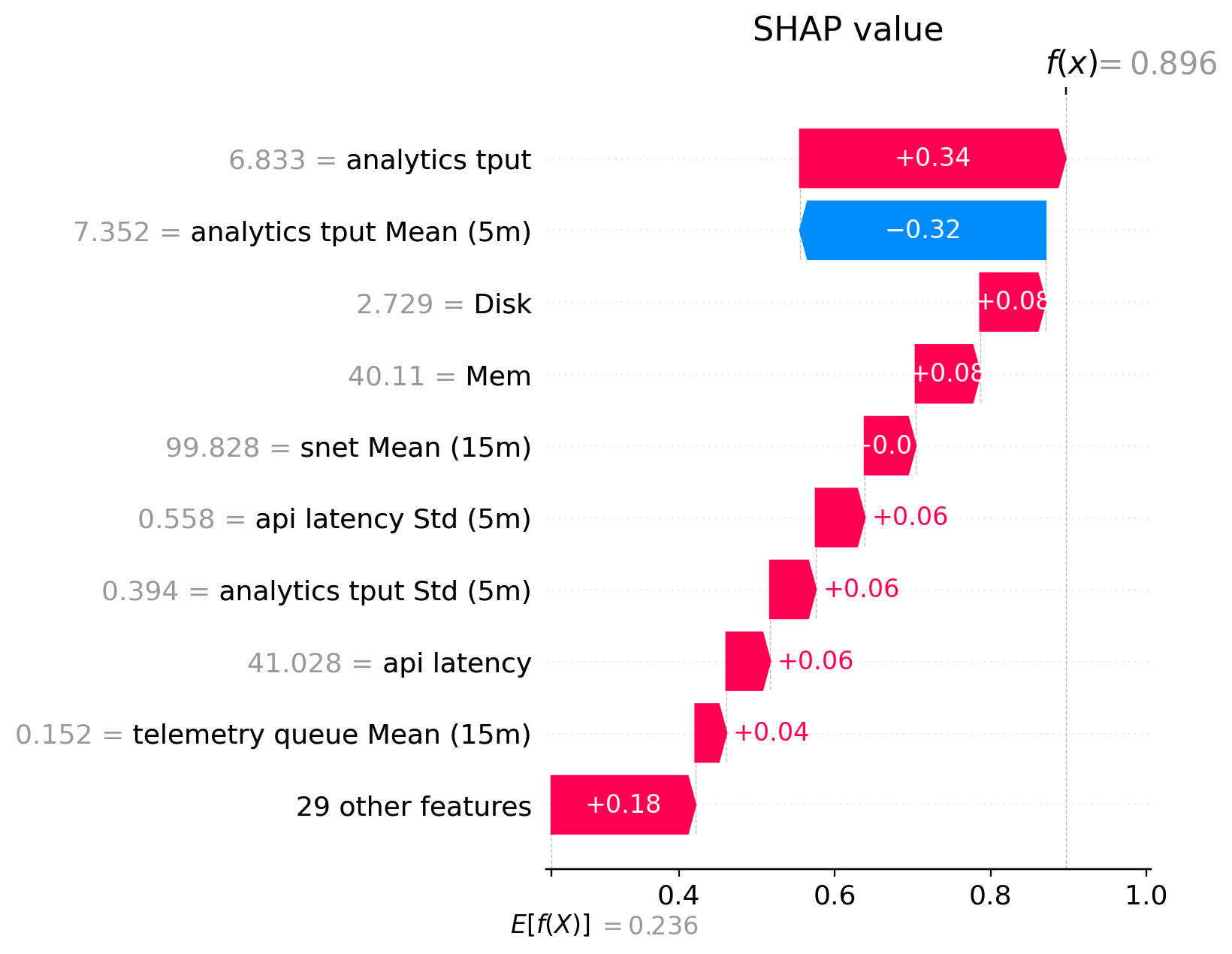}
    \caption{SHAP explanation for an analytics intent failure alert on the SDN-based edge-to-cloud testbed.}
    \label{fig:shap}
\end{figure}

To provide KPI-level operational insight, every alert from MILD is accompanied by a SHAP explanation that identifies how individual KPIs contribute to the predicted risk score. Fig.~\ref{fig:shap} shows the explanation for an analytics intent failure alert generated on the SDN-based edge-to-cloud testbed. The predicted raw risk score reaches 0.896 and its smoothed score exceeds the tuned decision threshold. The prediction is driven primarily by the current analytics throughput (\textit{analytics tput}), while current memory utilization (\textit{Mem}) and disk utilization (\textit{Disk}) provide additional positive evidence. In contrast, the 5-minute rolling mean of analytics throughput (\textit{analytics tput mean 5}) contributes negatively, indicating that the model contrasts the current workload with its recent historical behavior rather than reacting to a single instantaneous KPI measurement. Additional contributions from other engineered temporal features further refine the prediction. Overall, the explanation suggests an emerging analytics workload imbalance accompanied by increasing resource utilization, providing operators with a concise rationale for the alert and clear guidance for root-cause investigation.

\subsubsection{Dynamic Failure Urgency Estimation with Multi-Horizon Models}
The practical deployment of the multi-horizon framework, conceptually illustrated earlier in Fig. \ref{fig:multi_horizon}, provides progressively refined estimates of failure urgency. To validate this operational intelligence, we evaluated an ensemble of three models trained with prediction horizons of 120, 60, and 20 minutes on the statistical benchmark. During deployment, their first alerts for the failure of the Telemetry intent occurred approximately 128, 97, and 42 minutes before the observed failure, respectively. These values represent the empirical lead times achieved in practice, rather than the nominal training horizons. As additional horizon-specific models become active, the estimated failure window becomes narrower, enabling operators to transition from early planning to increasingly urgent mitigation actions.

\subsection{Ablation Study and Hyperparameter Sensitivity Analysis}
\label{subsec:ablation}
To rigorously validate the contribution of each component of
MILD's hybrid loss and teacher-augmented architecture, we conduct
a systematic ablation study and hyperparameter sensitivity analysis across \emph{all three} evaluation environments: the statistical benchmark, the microservices emulation testbed, and the SDN-based edge-to-cloud testbed. Five ablated variants are evaluated, each obtained by disabling exactly one component while holding all other hyperparameters at their chosen values. For sensitivity analysis, $\alpha$ and $w_c$, the two most influential scalar hyperparameters, are swept independently across five values while all other settings remain fixed. The four evaluation metrics are: average Failure Detection Rate
(\textbf{FDR}, \%), average Lead Time (\textbf{LT}, min),
False Positives per Day (\textbf{FP}), and Root-Cause
Disambiguation Accuracy (\textbf{DA}, \%).
The full MILD model appears in every table as the reference row.

\subsubsection{Ablation Result Analysis}
%----- Table 1: Ablation — statistical -----%
\begin{table}[!t]
\centering
\caption{Ablation on the Statistical Benchmark (10-fold blocked CV).
Bold = best per column; \textcolor{red!70!black}{red} = Worst Per Column.}
\label{tab:ablation_syn}
\scriptsize
\renewcommand{\arraystretch}{1.2}
\begin{tabular}{@{}lcccc@{}}
\toprule
\textbf{Variant} & \textbf{FDR (\%)} & \textbf{LT (min)} & \textbf{FP/day} & \textbf{DA (\%)} \\ \midrule
\textbf{MILD (full, proposed)}                          & 98.89 & 100.09 & 4.97 & 89.67 \\
w/o Distillation ($\alpha{=}1$)                        & 96.11 & \best{100.47} & 5.13 & 87.04 \\
w/o Gate Cause Sup.\ ($w_c{=}0$)                       & 98.89 & 100.31 & \worst{6.49} & 87.61 \\
w/o Teacher Aug.\ in Gate                              & \best{100.0}  & 98.27  & \best{4.13} & \textbf{89.71} \\
w/o Teacher KL in Gate   & \worst{91.11}  & \worst{93.83} & 4.78 & \worst{79.22} \\
w/o Decorrelation ($\lambda_{\mathrm{decorr}}{=}0$)   & \best{100.0}  & 100.04 & 5.77 & 88.67 \\
\bottomrule
\end{tabular}
\end{table}

%----- Table 2: Ablation — Microservices -----%
\begin{table}[!t]
\centering
\caption{Ablation on the Microservices Emulation Testbed (3-fold blocked CV).
Bold = best per column; \textcolor{red!70!black}{red} = Worst Per Column.}
\label{tab:ablation_ms}
\scriptsize
\renewcommand{\arraystretch}{1.2}
\begin{tabular}{@{}lcccc@{}}
\toprule
\textbf{Variant} & \textbf{FDR (\%)} & \textbf{LT (min)} & \textbf{FP/day} & \textbf{DA (\%)} \\ \midrule
\textbf{MILD (full, proposed)}                          & 91.32 & 31.76 & 8.17  & 66.80 \\
w/o Distillation ($\alpha{=}1$)                        & 91.57 & \best{32.84} & 11.96 & 67.35 \\
w/o Gate Cause Sup.\ ($w_c{=}0$)                       & 91.50 & 32.65 & \worst{19.39} & \worst{20.65} \\
w/o Teacher Aug.\ in Gate                              & \textbf{91.65} & 32.37 & 8.72  & \textbf{69.34} \\
w/o Teacher KL in Gate   & \worst{90.57}  & \worst{30.6} & \textbf{7.52} & 65.93 \\
w/o Decorrelation ($\lambda_{\mathrm{decorr}}{=}0$)   & 90.87 & 31.66 & 7.95  & 65.76 \\
\bottomrule
\end{tabular}
\end{table}

%----- Table 3: Ablation — Ryu -----%
\begin{table}[!t]
\centering
\caption{Ablation on the SDN-based Edge-to-Cloud Testbed (3-fold blocked CV).
Bold = best per column; \textcolor{red!70!black}{red} = Worst Per Column.}
\label{tab:ablation_ryu}
\scriptsize
\renewcommand{\arraystretch}{1.2}
\begin{tabular}{@{}lcccc@{}}
\toprule
\textbf{Variant} & \textbf{FDR (\%)} & \textbf{LT (min)} & \textbf{FP/day} & \textbf{DA (\%)} \\ \midrule
\textbf{MILD (full, proposed)}                          & 92.06 & \best{8.62} & 8.30 & \best{88.97} \\
w/o Distillation ($\alpha{=}1$)                        & 80.20 & 8.51 & 8.74 & 87.18 \\
w/o Gate Cause Sup.\ ($w_c{=}0$)                       & \best{95.64} & 8.52 & \worst{8.88} & 87.95 \\
w/o Teacher Aug.\ in Gate                              & 76.34 & 7.67 & 7.68 & 81.28 \\
w/o Teacher KL in Gate   & \worst{59.21}  & \worst{7.19} & \textbf{6.34} & \worst{67.44} \\
w/o Decorrelation ($\lambda_{\mathrm{decorr}}{=}0$)   & 90.78 & 8.23 & 8.06 & 88.72 \\
\bottomrule
\end{tabular}
\end{table}

Tables~\ref{tab:ablation_syn}--\ref{tab:ablation_ryu} report the ablation results. No single ablated variant dominates across all datasets; however, the full MILD model is the only configuration that avoids catastrophic failure on any single metric across all three environments, which is a critical requirement for real-world deployment.
\paragraph{Gate Cause Supervision \& Teacher Guidance} Gate cause supervision ($w_c$) is the most critical component for root-cause disambiguation. Removing it ($w_c=0$) collapses DA from 66.80\% to 20.65\% on the microservices testbed while more than doubling the false-positive rate, as the gate degenerates into an unguided correlation router. Similarly, removing the teacher KL term severely impairs both detection and attribution under noisy conditions, reducing the FDR to 59.21\% on the SDN-based edge-to-cloud testbed. While removing teacher augmentation slightly improves some metrics on the statistical benchmark, it severely degrades both detection and disambiguation under severe network impairments.
\paragraph{Distillation \& Decorrelation} Removing knowledge distillation ($\alpha=1$) has mixed effects in cleaner environments but causes a 12.15 percentage-point FDR drop on the SDN-based edge-to-cloud testbed that highlights its role as a robustness mechanism. Finally, head decorrelation ($\mathcal{L}_{\mathrm{decorr}}$) acts as a lightweight stabilizer. Although its removal does not cause catastrophic degradation, it consistently reduces DA across all datasets, supporting expert diversity.

\subsubsection{Hyperparameter Sensitivity Result Analysis}
%-----Sensitivity — statistical -----%
\begin{table}[!t]
\centering
\caption{Hyperparameter sensitivity on the Statistical Benchmark.
Bold = chosen configuration.}
\label{tab:sens_syn}
\scriptsize
\renewcommand{\arraystretch}{1.2}
\begin{tabular}{@{}clcccc@{}}
\toprule
\textbf{Param} & \textbf{Value} & \textbf{FDR (\%)} & \textbf{LT (min)} & \textbf{FP/day} & \textbf{DA (\%)} \\ \midrule
\multirow{5}{*}{$\alpha$}
 & 0.3               & 99.44 & 92.60  & 7.68  & 89.67 \\
 & 0.5               & 98.33 & 94.31  & 6.65   & 89.67 \\
 & 0.7               & 98.33 & 96.06  & 6.92  & 89.71 \\
 & \textbf{0.9}      & \textbf{98.89} & \textbf{100.09} & \textbf{4.97} & \textbf{89.67} \\
 & 1.0 (no distill.) & 96.11 & 100.47 & 5.13  & 87.04 \\ \midrule
\multirow{5}{*}{$w_c$}
 & 0.0 (no Gate Sup.) & 98.89 & 100.31 & 6.49  & 87.61 \\
 & 0.3                & 98.89 & 100.49 & 6.11   & 89.09 \\
 & 0.5                & 98.89 & 100.36 & 6.18   & 92.05 \\
 & \textbf{0.7}       & \textbf{98.89} & \textbf{100.09} & \textbf{4.97} & \textbf{89.67} \\
 & 0.9                & 98.33 & 101.00 & 6.24  & 91.38 \\
\bottomrule
\end{tabular}
\end{table}

%-----Sensitivity — Microservices -----%
\begin{table}[!t]
\centering
\caption{Hyperparameter sensitivity on the Microservices Testbed.
Bold = chosen configuration.}
\label{tab:sens_ms}
\scriptsize
\renewcommand{\arraystretch}{1.2}
\begin{tabular}{@{}clcccc@{}}
\toprule
\textbf{Param} & \textbf{Value} & \textbf{FDR (\%)} & \textbf{LT (min)} & \textbf{FP/day} & \textbf{DA (\%)} \\ \midrule
\multirow{6}{*}{$\alpha$}
 & 0.3               & 75.98 & 25.69 & 7.25  & 56.66 \\
 & 0.5               & 90.21 & 29.43 & 6.79  & 66.74 \\
 & \textbf{0.6}      & \textbf{91.32} & \textbf{31.76} & \textbf{8.17}  & \textbf{66.80} \\
 & 0.7               & 91.83 & 32.82 & 10.32 & 68.31 \\
 & 0.9      & 91.43 & 32.94 & 12.47 & 65.10 \\
 & 1.0 (no distill.) & 91.57 & 32.84 & 11.96 & 67.35 \\ \midrule
\multirow{6}{*}{$w_c$}
 & 0.0 (no Gate Sup.) & 91.50 & 32.65 & 19.39 & 20.65 \\
 & 0.3                & 91.57 & 32.00 & 10.89 & 64.27 \\
 & 0.5                & 90.58 & 31.63 &  7.01 & 65.95 \\
 & \textbf{0.6}       & \textbf{91.32} & \textbf{31.76} & \textbf{8.17}  & \textbf{66.80} \\
 & 0.7       & 90.4 & 31.48 & 9.33 & 64.34 \\
 & 0.9                & 91.07 & 31.82 &  9.70 & 68.18 \\
\bottomrule
\end{tabular}
\end{table}

%-----Sensitivity — Ryu -----%
\begin{table}[!t]
\centering
\caption{Hyperparameter sensitivity on the SDN-based Testbed.
Bold = chosen configuration.}
\label{tab:sens_ryu}
\scriptsize
\renewcommand{\arraystretch}{1.2}
\begin{tabular}{@{}clcccc@{}}
\toprule
\textbf{Param} & \textbf{Value} & \textbf{FDR (\%)} & \textbf{LT (min)} & \textbf{FP/day} & \textbf{DA (\%)} \\ \midrule
\multirow{6}{*}{$\alpha$}
 & 0.3               & 90.46 & 7.56 &  8.54 & 85.64 \\
 & 0.5               & 89.64 & 8.11 &  7.97 & 87.18 \\
 & 0.6      & 92.35 & 8.38 & 8.59 & 88.72 \\
 & \textbf{0.7}               & \textbf{92.06} & \textbf{8.62} &  \textbf{8.30} & \textbf{88.97} \\
 & 0.9               & 93.73 & 9.37 & 10.08 & 89.49 \\
 & 1.0 (no distill.) & 80.20 & 8.51 &  8.74 & 87.18 \\ \midrule
\multirow{6}{*}{$w_c$}
 & 0.0 (no Gate Sup.) & 95.64 & 8.52 & 8.88 & 87.95 \\
 & 0.3                & 93.34 & 8.37 & 8.35 & 87.95 \\
 & 0.5                & 91.63 & 8.27 & 8.35 & 88.21 \\
 & \textbf{0.6}       & \textbf{92.06} & \textbf{8.62} &  \textbf{8.30} & \textbf{88.97} \\
 & 0.7                & 90.39 & 8.19 & 8.30 & 87.69 \\
 & 0.9                & 90.78 & 8.13 & 8.11 & 89.23 \\
\bottomrule
\end{tabular}
\end{table}
Tables~\ref{tab:sens_syn}--\ref{tab:sens_ryu} report the model's sensitivity to the distillation mixing coefficient ($\alpha$) and the gate cause supervision weight ($w_c$). Because no single configuration is uniformly optimal across all environments, the chosen operating points (bolded) reflect the best dataset-specific compromise between detection sensitivity, attribution accuracy, and false-positive control.
\paragraph{Distillation Coefficient ($\alpha$)} Balancing focal loss and teacher KL divergence, moderate-to-high $\alpha$ values generally yield the best performance. While $\alpha{=}1.0$ (no distillation) slightly extends lead times on the statistical benchmark, it increases FP/day and reduces Disambiguation Accuracy (DA) in more challenging environments. We selected $\alpha{=}0.9$ for the controlled benchmark, and more conservative values of $\alpha{=}0.6$ and $\alpha{=}0.7$ for the microservices and SDN-based edge-to-cloud testbeds, respectively, reflecting their different noise characteristics and preventing FP inflation (e.g., $\alpha{=}0.7$ increases FP/day by approximately 26\% on the microservices testbed).

\paragraph{Gate Cause Supervision ($w_c$)} This parameter dictates the strength of the causal alignment term $\mathrm{KL}(\mathbf{y}^{\mathrm{cause}}_t \| \mathbf{g}_t)$. Setting $w_c{=}0$ triggers a catastrophic DA collapse on the microservices testbed and degrades the overall operating profile on the SDN testbed, confirming the necessity of explicit causal supervision. However, larger supervision weights ($w_c \ge 0.9$) provide only marginal gains in DA without consistently improving the overall multi-metric profile. Consequently, moderate values ($w_c \in [0.6,0.7]$) are selected to maintain a balanced trade-off between detection performance, attribution quality, and false-positive control.

\vspace{-0.3cm}
\section{Discussion}
\label{sec:discussion}
\textbf{Real-World Deployment Pathway}: Deploying MILD in production requires a phased, human-in-the-loop strategy \cite{feamster2017and,princeton}. Initial pre-training on statistical benchmarks captures fundamental co-drift patterns, drastically reducing the annotation burden for subsequent fine-tuning on a small ``golden dataset'' of operational telemetry. Post-deployment, uncertain alerts (where $\max_i g_{t,i}$ is below a confidence threshold) are surfaced for operator verification, driving continuous, incremental retraining.

\textbf{Generalizability Beyond the Three-Intent Use Case}: while evaluated on three macro-intents, MILD's architecture scales naturally. Accommodating additional intents requires only appending expert-head pairs and expanding the gate's output dimension, making the framework readily adaptable to broader IBN environments, such as full-stack 5G slice assurance or enterprise networks.

\textbf{Limitations of MILD}: Despite its strong performance, MILD's supervised nature relies on post-mortem expert annotation of failure events. Furthermore, the model currently assumes a static root-cause intent per failure window. Extending this to track dynamic, cascading root causes during prolonged incidents remains an open challenge.
\section{Conclusion}
\label{sec:conclusion}
This paper introduced MILD, a framework that reformulates intent assurance from reactive drift detection to proactive intent failure prediction. By modeling the self-driving network control loop as a causally linked multi-intent system, we demonstrated how single faults trigger ambiguous co-drift anomalies.
%MILD resolves this through a teacher-augmented Mixture-of-Experts architecture optimized for joint failure prediction and root-cause attribution, supported by explainability and multi-horizon urgency estimation.
MILD resolves this through a teacher-augmented Mixture-of-Experts architecture optimized for joint failure prediction and root-cause attribution. MILD enables both intent-level root-cause disambiguation and KPI-level diagnostic explanations, as well as multi-horizon modeling for failure urgency estimation. Validated across the statistical benchmark, microservices, and SDN-based edge-to-cloud emulation testbeds, MILD consistently delivered high detection rates, substantial operational lead times, and root-cause attribution, remaining robust even under severe network impairments in the SDN-based testbed. Future work will focus on data-efficient semi-supervised and active learning, validation on production telemetry, and dynamic causal modeling to track shifting root causes in complex, multi-stage incidents.
\bibliographystyle{IEEEtran}
\bibliography{references}

@INPROCEEDINGS{10770652,
  author={Muonagor, Chukwuemeka and Bensalem, Mounir and Jukan, Admela},
  booktitle={Proc. IEEE LATINCOM}, 
  title={Performance Analysis of Learning-based Intent Drift Detection Algorithms in Next Generation Networks}, 
  year={2024},
  volume={},
  number={},
  pages={1-6},
  doi={10.1109/LATINCOM62985.2024.10770652}}

@misc{hossain2026mildmultiintentlearningdisambiguation,
      title={MILD: Multi-Intent Learning and Disambiguation for Proactive Failure Prediction in Intent-based Networking}, 
      author={Md. Kamrul Hossain and Walid Aljoby},
      year={2026},
      eprint={2602.14283},
      archivePrefix={arXiv},
      primaryClass={cs.NI},
      url={https://arxiv.org/abs/2602.14283}, 
}

@ARTICLE{7563819,
  author={Chen, Pengfei and Qi, Yong and Hou, Di},
  journal={IEEE Transactions on Services Computing}, 
  title={CauseInfer: Automated End-to-End Performance Diagnosis with Hierarchical Causality Graph in Cloud Environment}, 
  year={2019},
  volume={12},
  number={2},
  pages={214-230},
  doi={10.1109/TSC.2016.2607739}}

@ARTICLE{11229957,
  author={Zhang, Xiao and Wang, Qi and Li, Mingyi and Yuan, Yuan and Xiao, Mengbai and Zhuang, Fuzhen and Yu, Dongxiao},
  journal={IEEE Transactions on Services Computing}, 
  title={TAMO:Fine-Grained Root Cause Analysis via Tool-Assisted LLM Agent With Multi-Modality Observation Data in Cloud-Native Systems}, 
  year={2025},
  volume={18},
  number={6},
  pages={4221-4233},
  doi={10.1109/TSC.2025.3629066}}

@misc{tang2025microrcaagentmicroservicerootcause,
      title={MicroRCA-Agent: Microservice Root Cause Analysis Method Based on Large Language Model Agents}, 
      author={Pan Tang and Shixiang Tang and Huanqi Pu and Zhiqing Miao and Zhixing Wang},
      year={2025},
      eprint={2509.15635},
      archivePrefix={arXiv},
      primaryClass={cs.AI},
      url={https://arxiv.org/abs/2509.15635}, 
}

@INPROCEEDINGS{9110353,
  author={Wu, Li and Tordsson, Johan and Elmroth, Erik and Kao, Odej},
  booktitle={Proc. IEEE/IFIP NOMS}, 
  title={MicroRCA: Root Cause Localization of Performance Issues in Microservices}, 
  year={2020},
  volume={},
  number={},
  pages={1-9},
  doi={10.1109/NOMS47738.2020.9110353}}

@article{feamster2017and,
  title={Why (and how) networks should run themselves},
  author={Feamster, Nick and Rexford, Jennifer},
  journal={arXiv preprint arXiv:1710.11583},
  year={2017}
}

@inproceedings{cogswell2016reducing,
  author    = {Michael Cogswell and Faruk Ahmed and Ross Girshick and C. Lawrence Zitnick and Dhruv Batra},
  title     = {Reducing Overfitting in Deep Networks by Decorrelating Representations},
  booktitle = {4th International Conference on Learning Representations (ICLR 2016)},
  address   = {San Juan, Puerto Rico},
  month     = may,
  year      = {2016}
}

@ARTICLE{11334180,
  author={Gharbaoui, Molka and Sciarrone, Filippo and Fontana, Mattia and Castoldi, Piero and Martini, Barbara},
  journal={IEEE Transactions on Network and Service Management}, 
  title={Assurance and Conflict Detection in Intent-Based Networking: A Comprehensive Survey and Insights on Standards and Open-Source Tools}, 
  year={2026},
  volume={23},
  number={},
  pages={1891-1912},
  doi={10.1109/TNSM.2026.3651896}}

@misc{princeton,
	author = {Feamster and Rexford},
	title = {Workshop on Self-Driving Networks — Workshop Report},
	howpublished = {\url{https://www.cs.princeton.edu/~jrex/papers/self-driving-networks18.pdf}},
    institution = {Princeton University},
	year = {2018},
	note = {[Accessed 07-06-2026]},
}

@misc{hossain2026leaddriftrealtimeexplainableintent,
      title={LEAD-Drift: Real-time and Explainable Intent Drift Detection by Learning a Data-Driven Risk Score}, 
      author={Md. Kamrul Hossain and Walid Aljoby},
      year={2026},
      eprint={2602.13672},
      archivePrefix={arXiv},
      primaryClass={cs.NI},
      url={https://arxiv.org/abs/2602.13672}, 
}

@INPROCEEDINGS{10942926,
  author={Gharbaoui, M. and Martini, B. and Berardi, D. and Castoldi, P.},
  booktitle={Proc. IEEE ICIN}, 
  title={Towards Intent Assurance: A Traffic Prediction Model for Software-Defined Networks}, 
  year={2025},
  volume={},
  number={},
  pages={135-139},
  doi={10.1109/ICIN64016.2025.10942926}}

@book{Hyndman_Athanasopoulos_2021, place={Melbourne}, edition={3rd}, title={Forecasting: Principles and practice}, publisher={OTexts}, author={Hyndman, Rob J. and Athanasopoulos, George}, year={2021}}

@inproceedings{10.5555/3295222.3295230,
author = {Lundberg, Scott M. and Lee, Su-In},
title = {A unified approach to interpreting model predictions},
year = {2017},
isbn = {9781510860964},
publisher = {Curran Associates Inc.},
address = {Red Hook, NY, USA},
booktitle = {Proc. NIPS},
pages = {4768–4777},
numpages = {10},
location = {Long Beach, California, USA},
series = {NIPS'17}
}

@INPROCEEDINGS{9557387,
  author={Züfle, Marwin and Agne, Joachim and Grohmann, Johannes and Dörtoluk, Ibrahim and Kounev, Samuel},
  booktitle={Proc. IEEE INDIN}, 
  title={A Predictive Maintenance Methodology: Predicting the Time-to-Failure of Machines in Industry 4.0}, 
  year={2021},
  volume={},
  number={},
  pages={1-8},
  doi={10.1109/INDIN45523.2021.9557387}}

@ARTICLE{11293797,
  author={Hossain, Md. Kamrul and Aljoby, Walid},
  journal={IEEE Open Journal of the Communications Society}, 
  title={NetIntent: Leveraging Large Language Models for End-to-End Intent-Based SDN Automation}, 
  year={2025},
  volume={6},
  number={},
  pages={10512-10541},
  doi={10.1109/OJCOMS.2025.3642642}}

@article{Basikolo2023TowardsZD,
  title={Towards zero downtime: Using machine learning to predict network failure in 5G and beyond},
  author={Emmanuel Basikolo and Thomas Basikolo},
  journal={ITU Journal on Future and Evolving Technologies},
  volume={4},
  number={3},
  pages={434--446},
  year={2023},
  publisher={ITU},
  doi = "10.52953/PYAF8065"
}

@article{10.1145/3501297,
author = {Soldani, Jacopo and Brogi, Antonio},
title = {Anomaly Detection and Failure Root Cause Analysis in (Micro) Service-Based Cloud Applications: A Survey},
year = {2022},
issue_date = {March 2023},
publisher = {Association for Computing Machinery},
address = {New York, NY, USA},
volume = {55},
number = {3},
issn = {0360-0300},
url = {https://doi.org/10.1145/3501297},
doi = {10.1145/3501297},
journal = {ACM Comput. Surv.},
month = feb,
articleno = {59},
numpages = {39}
}

@misc{tarekegn2024deeplearningmultilabellearning,
      title={Deep Learning for Multi-Label Learning: A Comprehensive Survey}, 
      author={Adane Nega Tarekegn and Mohib Ullah and Faouzi Alaya Cheikh},
      year={2024},
      eprint={2401.16549},
      archivePrefix={arXiv},
      primaryClass={cs.LG},
      url={https://arxiv.org/abs/2401.16549}, 
}

@article{hinton2015distilling,
  title={Distilling the knowledge in a neural network},
  author={Hinton, Geoffrey and Vinyals, Oriol and Dean, Jeff},
  journal={arXiv preprint arXiv:1503.02531},
  year={2015}
}

@inproceedings{lin2017focal,
  title={Focal loss for dense object detection},
  author={Lin, Tsung-Yi and Goyal, Priya and Girshick, Ross and He, Kaiming and Doll{\'a}r, Piotr},
  booktitle={Proc. ICCV},
  pages={2980--2988},
  year={2017}
}

@article{chen2022towards,
  title={Towards understanding the mixture-of-experts layer in deep learning},
  author={Chen, Zixiang and Deng, Yihe and Wu, Yue and Gu, Quanquan and Li, Yuanzhi},
  journal={Advances in neural information processing systems},
  volume={35},
  pages={23049--23062},
  year={2022}
}

@article{LIU2026111857,
title = {Refined link-level intent drift forecasting through advanced link performance prediction and path similarity approaches},
journal = {Computer Networks},
volume = {275},
pages = {111857},
year = {2026},
issn = {1389-1286},
doi = {https://doi.org/10.1016/j.comnet.2025.111857},
author = {Hanlin Liu and Hua Li and Yintan Ai},
}

@article{DEHGHANBIYAR2025111561,
title = {Autonomous conflict handling in intent-based management},
journal = {Computer Networks},
volume = {271},
pages = {111561},
year = {2025},
issn = {1389-1286},
doi = {https://doi.org/10.1016/j.comnet.2025.111561},
url = {https://www.sciencedirect.com/science/article/pii/S1389128625005286},
author = {Elham {Dehghan Biyar} and Mirko D’Angelo and Josué Castañeda Cisneros and Amadeu Nascimento and Marin Orlic and Ankita Likhyani and András Zahemszky and Ahmet Cihat Baktir and Dagnachew Azene Temesgene and Dinand Roeland},
}

@article{VIOLOS2026111872,
title = {Detecting application transitions and identifying application types for intent-based network assurance: A machine learning perspective},
journal = {Computer Networks},
volume = {274},
pages = {111872},
year = {2026},
issn = {1389-1286},
doi = {https://doi.org/10.1016/j.comnet.2025.111872},
url = {https://www.sciencedirect.com/science/article/pii/S1389128625008382},
author = {John Violos and Fotios Voutsas and Christos Diou and Aris Leivadeas},
}

@ARTICLE{9592636,
  author={Jalodia, Nikita and Taneja, Mohit and Davy, Alan},
  journal={IEEE Open Journal of the Communications Society}, 
  title={A Deep Neural Network-Based Multi-Label Classifier for SLA Violation Prediction in a Latency Sensitive NFV Application}, 
  year={2021},
  volume={2},
  number={},
  pages={2469-2493},
  doi={10.1109/OJCOMS.2021.3122844}}

@article{Zanouda_2024,
title={Telecom Foundation Models: Applications, Challenges, and Future Trends},
url={http://dx.doi.org/10.36227/techrxiv.172296173.35282215/v1},
DOI={10.36227/techrxiv.172296173.35282215/v1},
publisher={Institute of Electrical and Electronics Engineers (IEEE)},
author={Zanouda, Tahar and Masoudi, Meysam and Gebre, Fitsum Gaim and Dohler, Mischa},
year={2024},
month=aug }

@ARTICLE{10529727,
  author={Kaushik, Aryan and Singh, Rohit and Dayarathna, Shalanika and Senanayake, Rajitha and Di Renzo, Marco and Dajer, Miguel and Ji, Hyoungju and Kim, Younsun and Sciancalepore, Vincenzo and Zappone, Alessio and Shin, Wonjae},
  journal={IEEE Communications Standards Magazine}, 
  title={Toward Integrated Sensing and Communications for 6G: Key Enabling Technologies, Standardization, and Challenges}, 
  year={2024},
  volume={8},
  number={2},
  pages={52-59},
  doi={10.1109/MCOMSTD.0007.2300043}}

@article{d2023orchestran,
  title={OrchestRAN: Orchestrating network intelligence in the open RAN},
  author={D’Oro, Salvatore and Bonati, Leonardo and Polese, Michele and Melodia, Tommaso},
  journal={IEEE Transactions on Mobile Computing},
  volume={23},
  number={7},
  pages={7952--7968},
  year={2023},
  publisher={IEEE}
}

@misc{ardestani2025nwdafenabledanalyticsclosedloopautomation,
      title={Towards NWDAF-enabled Analytics and Closed-Loop Automation in 5G Networks}, 
      author={Fatemeh Shafiei Ardestani and Niloy Saha and Noura Limam and Raouf Boutaba},
      year={2025},
      eprint={2505.06789},
      archivePrefix={arXiv},
      primaryClass={cs.NI},
      url={https://arxiv.org/abs/2505.06789}, 
}

@inproceedings{10.1145/3580305.3599934,
author = {Wang, Lu and Zhang, Chaoyun and Ding, Ruomeng and Xu, Yong and Chen, Qihang and Zou, Wentao and Chen, Qingjun and Zhang, Meng and Gao, Xuedong and Fan, Hao and Rajmohan, Saravan and Lin, Qingwei and Zhang, Dongmei},
title = {Root Cause Analysis for Microservice Systems via Hierarchical Reinforcement Learning from Human Feedback},
year = {2023},
isbn = {9798400701030},
publisher = {Association for Computing Machinery},
address = {New York, NY, USA},
url = {https://doi.org/10.1145/3580305.3599934},
doi = {10.1145/3580305.3599934},
booktitle = {Proc. ACM SIGKDD},
pages = {5116–5125},
numpages = {10},
location = {Long Beach, CA, USA},
series = {KDD '23}
}

@ARTICLE{10456766,
  author={Arellano-Uson, Jesus and Magaña, Eduardo and Morató, Daniel and Izal, Mikel},
  journal={IEEE Access}, 
  title={Interactivity Anomaly Detection in Remote Work Scenarios Using LSTM}, 
  year={2024},
  volume={12},
  number={},
  pages={34402-34416},
  doi={10.1109/ACCESS.2024.3372405}}

@misc{rfc9315,
    series =    {Request for Comments},
    number =    9315,
    howpublished =  {RFC 9315},
    publisher = {RFC Editor},
    doi =       {10.17487/RFC9315},
    url =       {https://www.rfc-editor.org/info/rfc9315},
    author =    {Alexander Clemm and Laurent Ciavaglia and Lisandro Zambenedetti Granville and Jeff Tantsura},
    title =     {{Intent-Based Networking - Concepts and Definitions}},
    pagetotal = 23,
    year =      2022,
    month =     oct
}

@misc{opendaylight2015,
  author       = {OpenDaylight Project},
  title        = {OpenDaylight: A Linux Foundation Collaborative Project},
  year         = {2015},
  url          = {https://www.opendaylight.org/},
  note         = {Accessed: 2025-01-15}
}

@INPROCEEDINGS{10001426,
  author={Perepu, Satheesh K. and Martins, Jean P. and S, Ricardo Souza and Dey, Kaushik},
  booktitle={Proc. IEEE GLOBECOM}, 
  title={Intent-based multi-agent reinforcement learning for service assurance in cellular networks}, 
  year={2022},
  volume={},
  number={},
  pages={2879-2884},
  doi={10.1109/GLOBECOM48099.2022.10001426}}

@INPROCEEDINGS{9615580,
  author={Zheng, Xiaoang and Leivadeas, Aris},
  booktitle={Proc. IEEE CNSM}, 
  title={Network Assurance in Intent-Based Networking Data Centers with Machine Learning Techniques}, 
  year={2021},
  volume={},
  number={},
  pages={14-20},
  doi={10.23919/CNSM52442.2021.9615580}}

@article{sharma2023comprehensive,
  title={A Comprehensive Survey on Network Resource Management in SDN Enabled Data Centre Network},
  author={Sharma, Ashish and Tokekar, Sanjiv and Varma, Sunita},
  journal={6G Enabled Fog Computing in IoT: Applications and Opportunities},
  pages={333--353},
  year={2023},
  publisher={Springer}
}

@misc{HossainMILD, title={MILD: Multi-Intent Learning and Disambiguation for Proactive Failure Prediction in Intent-based Networking}, 
url={https://github.com/Muhammadkamrul/Extended_MILD}, 
journal={GitHub}, 
author={Hossain, Md. Kamrul and Aljoby, Walid}
}

@ARTICLE{9925251,
  author={Leivadeas, Aris and Falkner, Matthias},
  journal={IEEE Communications Surveys \& Tutorials}, 
  title={A Survey on Intent-Based Networking}, 
  year={2023},
  volume={25},
  number={1},
  pages={625-655},
  doi={10.1109/COMST.2022.3215919}}

@INPROCEEDINGS{11073595,
  author={Dzeparoska, Kristina and Leon-Garcia, Alberto},
  booktitle={Proc. IEEE/IFIP NOMS}, 
  title={KPI Assurance and LLMs for Intent-Based Management}, 
  year={2025},
  volume={},
  number={},
  pages={1-9},
  doi={10.1109/NOMS57970.2025.11073595}}

@INPROCEEDINGS{10575429,
  author={Dzeparoska, Kristina and Tizghadam, Ali and Leon-Garcia, Alberto},
  booktitle={Proc. IEEE/IFIP NOMS}, 
  title={Intent Assurance using LLMs guided by Intent Drift}, 
  year={2024},
  volume={},
  number={},
  pages={1-7},
  doi={10.1109/NOMS59830.2024.10575429}}

@INPROCEEDINGS{10620891,
  author={Ouyang, Ying and Li, Changle and Zhang, Jingwen and Zhao, Xiaoxue and Yang, Chungang},
  booktitle={Proc. IEEE INFOCOM Workshop}, 
  title={Intent-Driven 6G End-to-End Network Orchestration}, 
  year={2024},
  volume={},
  number={},
  pages={1-2},
  doi={10.1109/INFOCOMWKSHPS61880.2024.10620891}}
\end{document}